\documentclass[aps,prb,amsmath,amssymb,floatfix,twocolumn,showpacs,10pt]{revtex4-1}
\usepackage{times}
\usepackage{bbold}
\usepackage{bm}
\usepackage{graphicx} 
\usepackage{color}
\usepackage{hyperref}
\usepackage{textcomp}
\usepackage{amssymb} 
\usepackage{wasysym}

\begin{document}

\title{Antiferromagnetic nuclear spin helix and topological superconductivity in \texorpdfstring{$^{13}$}~C nanotubes}

\author{Chen-Hsuan Hsu$^{1}$}
\author{Peter Stano$^{1,2}$}
\author{Jelena Klinovaja$^{3}$}
\author{Daniel Loss$^{1,3}$}

\affiliation{$^{1}$RIKEN Center for Emergent Matter Science (CEMS), Wako, Saitama 351-0198, Japan}
\affiliation{$^{2}$Institute of Physics, Slovak Academy of Sciences, 845 11 Bratislava, Slovakia}
\affiliation{$^{3}$Department of Physics, University of Basel, Klingelbergstrasse 82, CH-4056 Basel, Switzerland}

\date{\today}

\begin{abstract}

We investigate the Ruderman-Kittel-Kasuya-Yosida (RKKY) interaction arising from the hyperfine coupling between localized nuclear spins and conduction electrons in interacting $^{13}$C carbon nanotubes. Using the Luttinger liquid formalism, we show that the RKKY interaction is sublattice dependent, consistent with the spin susceptibility calculation in noninteracting carbon nanotubes, and it leads to an antiferromagnetic nuclear spin helix in finite-size systems. The transition temperature reaches up to tens of mK, due to a strong boost by a positive feedback through the Overhauser field from ordered nuclear spins. Similar to GaAs nanowires, the formation of the helical nuclear spin order gaps out half of the conduction electrons, and is therefore observable as a reduction of conductance by a factor of 2 in a transport experiment.
The nuclear spin helix leads to a density wave combining spin and charge degrees of freedom in the electron subsystem, resulting in synthetic spin-orbit interaction, which induces non-trivial topological phases. As a result, topological superconductivity with Majorana fermion bound states can be realized in the system in the presence of proximity-induced superconductivity without the need of fine tuning the chemical potential. 
We present the phase diagram as a function of system parameters, including the pairing gaps, the gap due to the nuclear spin helix, and the Zeeman field perpendicular to the helical plane.

\end{abstract}
 
 
\maketitle

\section{Introduction}

The search for topological superconductivity and exotic quasiparticles supported by it, such as Majorana fermions (MFs), remains an ongoing challenge. MFs, being their own antiparticles, exhibit non-Abelian statistics and are promising candidates for realization of quantum computation.~\cite{Nayak:2008,Alicea:2012} Despite intensive experimental~\cite{Mourik:2012,Das:2012,Deng:2012,Finck:2013,Nadj-Perge:2014,Pawlak:2015} and theoretical~\cite{Hasan:2010,Qi:2011,Fu:2008,Lutchyn:2010,Oreg:2010} efforts, the observation of MFs still remains inconclusive. It is therefore important to propose experimentally achievable devices for the realization of such particles and even more exotic parafermions.~\cite{Lindner:2012,Clarke:2013,Klinovaja:2014a,Klinovaja:2014b,Klinovaja:2014c,Mong:2014} 
In this respect, carbon nanotubes (CNTs) seem promising due to advantages such as the availability of high-quality samples, high tunability, and strong electron-electron interactions due to the spatial confinement,~\cite{Bockrath:1997,Tans:1997,Laird:2014,Sanchez-Valencia:2014} which is crucial for fractional statistics.

In a nanowire with free carriers (electrons) and localized spins, such as spins of atomic nuclei, these two subsystems are coupled by the hyperfine interaction.~\cite{Fischer:2009,Yazyev:2008,Csiszar:2014} With parameters typical for semiconducting nanowires, this interaction is weak on the scale of the electronic Fermi energy. It can then be recast as the Ruderman-Kittel-Kasuya-Yosida (RKKY) exchange interaction,~\cite{Ruderman:1954,Kasuya:1956,Yosida:1957,Ziman:1972} the electron mediated pairwise interaction between localized spins (see also Ref.~\citenum{Schecter:2015} for systems beyond the RKKY picture). The strength of this pairwise interaction is given by the many-body state of the electron subsystem, and reflects its properties. For example, the RKKY coupling as a function of distance is modulated at the electron Fermi wavelength,~\cite{Giuliani:2005} while the spin-orbit interaction of electrons results in spin anisotropies~\cite{Imamura:2004,Klinovaja:2013a} 
or suppression~\cite{Chesi:2010} of the RKKY coupling. In low dimensional systems, the effect of electron-electron interactions becomes striking.~\cite{Simon:2008,Braunecker:2009a,Braunecker:2009b} 
Namely, the RKKY interaction is strongly enhanced, formally seen as the renormalization of the exponent describing the power-law decay of correlators calculated in the Luttinger liquid formalism.~\cite{Egger:1996} The stronger the electron-electron interactions, the more pronounced is this enhancement around the electronic Fermi momentum, which leads to a sharp resonant peak in the RKKY coupling in one-dimensional systems. At low enough temperature, this peak results in the ordering of the localized spins into a helix, which corresponds to an effective rotating magnetic field (Overhauser field) seen by the electrons. This macroscopic field changes the electronic state by opening a partial gap at the Fermi energy. While this further boosts the RKKY coupling strength, it is also interesting on its own. Because a rotating field has a definite helicity, the partial gap opens in a spin selective way~\cite{Braunecker:2010,Meng:2013a} and the electronic subsystem also becomes helical. It has been theoretically suggested to exploit such helical Overhauser fields for, e.g., dynamical nuclear spin polarization,~\cite{Kornich:2015} stabilization of fractionalized fermions,~\cite{Klinovaja:2012b} or production of tune-free topological matter.~\cite{Klinovaja:2013b,Braunecker:2013,Vazifeh:2013} Signatures of such a partial gap opening have been observed in GaAs quantum wires in transport experiments at sub-K temperature.~\cite{Scheller:2014}
Subsequent density matrix renormalization-group analysis~\cite{Smerat:2011} also supported the formation of the RKKY-induced magnetic order discussed in Refs.~\citenum{Braunecker:2009a,Braunecker:2009b}.

In this paper we revisit the above picture considering metallic CNTs enriched by $^{13}$C, the atomic isotope with nuclear spin $\frac{1}{2}$.~\cite{Simon:2005,Rummeli:2007,Churchill:2009a,Churchill:2009b} While Refs.~\citenum{Braunecker:2009a,Braunecker:2009b} also considered CNTs, the presence of sublattices was omitted. On the other hand, the results for the spin susceptibility calculated in the noninteracting limit~\cite{Saremi:2007,Brey:2007,Black-Schaffer:2010,Kogan:2011,Klinovaja:2013a,Power:2013} suggest that the RKKY interaction is locally (between nearest neighbors) antiferromagnetic, unlike in, e.g., GaAs where it is ferromagnetic. The question therefore arises whether in the presence of strong electron-electron interactions, the RKKY interaction retains its locally antiferromagnetic character, and whether a macroscopic Overhauser field can still arise, which is necessary to push the transition temperature of the nuclear order to experimentally achievable values, and to offer $^{13}$C enriched CNT as a self-tuned topological matter platform.

To this end, we derive here the RKKY interaction taking into account sublattices explicitly in a CNT with interacting electrons. We find that the RKKY interaction is sublattice dependent, consistent with Refs.~\citenum{Saremi:2007,Brey:2007,Black-Schaffer:2010,Kogan:2011,Klinovaja:2013a,Power:2013}, and it leads to a locally {\it antiferromagnetic} nuclear spin helix.~\footnote[1]{A different (ferromagnetic) helical order was found in Ref.~\citenum{Lazic:2014}, which considered semiconducting noninteracting $^{13}$C CNTs, unlike metallic ones which we study here. The difference is expected, since in the former case the spin susceptibility depends weakly on the sublattices and is dominantly ferromagnetic, as discussed in Ref.~\citenum{Klinovaja:2013a}.} However, despite a lack of {\it macroscopic} spin polarization, the helix transition temperature is still strongly enhanced, and reaches several tens of mK.
We also confirm that the nuclear spin helix combining charge and spin degrees of freedom generates synthetic spin-orbit interaction for electrons, suitable to induce non-trivial topology~\cite{Hasan:2010,Qi:2011,Klinovaja:2013b} supporting MFs~\cite{Gangadharaiah:2011,Stoudenmire:2011} without involving intrinsic spin-orbit interactions that happen to be still weak in CNTs.~\cite{Izumida:2009,Klinovaja:2011a,Klinovaja:2011b} 
Therefore, we suggest to pursue experimentally the possibility to establish the RKKY induced nuclear spin order at low temperatures in CNTs highly enriched by $^{13}$C.

This paper is organized as follows. In Sec.~\ref{Sec:RKKY} we obtain the RKKY interaction within the Luttinger liquid formalism: in Sec.~\ref{SubSec:Hamiltonian},
we first establish the RKKY Hamiltonian in terms of the spin susceptibility of the conduction electrons, which are described as Luttinger liquid in Sec.~\ref{SubSec:Hel}; the bosonization of the electron spin operators, which enter the spin susceptibility, are discussed in Sec.~\ref{SubSec:spin}; finally, using the results of Sec.~\ref{SubSec:Hamiltonian}-\ref{SubSec:spin}, we obtain the RKKY interaction in Sec.~\ref{SubSec:chi}. 
The resulting Hamiltonian then represents a spin model for the nuclear spins, allowing us to investigate the nuclear spin order in Sec.~\ref{Sec:helical}: we take the ansatz for antiferromagnetic nuclear spin helix in Sec.~\ref{SubSec:magnon}, and compute the magnon (spin wave) spectrum; in Sec.~\ref{SubSec:Tc}, the transition temperature of the nuclear spin helix (without the feedback) is estimated. 
In Sec.~\ref{Sec:feedback}, we examine the feedback effect due to the nuclear Overhauser field: 
in Sec.~\ref{SubSec:Overhauser}, we show that the intervalley back scattering terms in the spin susceptibility are suppressed by the feedback; in Sec.~\ref{SubSec:RG}, we analyze the renormalized Overhauser field; in Sec.~\ref{SubSec:Tc_fb}, we estimate the transition temperature in the presence of the feedback, which is enhanced by more than four orders of magnitude.
In Sec.~\ref{Sec:RKKY_gapped} we investigate how the proximity-induced superconductivity affects the RKKY interaction: in Sec.~\ref{SubSec:chi_gapped}, we compute the spin susceptibility in the presence of the pairing gap; in Sec.~\ref{SubSec:T_gapped}, we show that the reduced transition temperature may still be within experimentally accessible regimes. 
In Sec.~\ref{Sec:Top_Su} we focus on the topological properties:
in Sec.~\ref{SubSec:refermionization}, a refermionized Hamiltonian is established, which allows us to find MF solutions straightforwardly; the  topological phase diagram is presented in Sec.~\ref{SubSec:MF}. 
Finally, we give a discussion on the nuclear spin helix and MFs in CNTs in Sec.~\ref{Sec:discussion}. The details of the calculations on the spin susceptibility in the presence of the pairing gap and solving the Schr\"odinger equation for MF solutions are given in Appendix~\ref{Sec:chi_gapped} and \ref{Sec:MF_solution}, respectively.

\section{RKKY interaction \label{Sec:RKKY}}
\subsection{Hyperfine and RKKY Hamiltonians \label{SubSec:Hamiltonian}}

Nuclear spins of $^{13}$C atoms embedded within CNTs couple to 
conduction electrons via the hyperfine exchange interaction. We consider a single-wall armchair-edged nanotube~\footnote[2]{Our calculation applies to any conducting CNTs. Here we focus on armchair-edged CNTs for the ease of notation.}
with the Hamiltonian
\begin{equation}
\mathcal{H}=\mathcal{H}_{\textrm{el}}+\mathcal{H}_{\textrm{hf}}.
\end{equation}
Here $\mathcal{H}_{\textrm{el}}$, discussed in Sec.~\ref{SubSec:Hel}, describes the interacting conduction electrons, and $\mathcal{H}_{\textrm{hf}}$ is the hyperfine coupling between the conduction electrons and localized nuclear spins. The dipolar interaction between the localized spins is much smaller than these two terms and hence neglected.~\cite{Paget:1977}
(for systems where this is not the case, the combined effect of direct and RKKY interactions may lead to, e.g., a canted spin state.~\cite{Costa:2005})

Assuming the electrons are in the lowest transverse mode due to a large transverse level spacing of the order of eV,
we obtain an effective one-dimensional hyperfine interaction,
\begin{equation}
\mathcal{H}_{\textrm{hf}} = \sum_{\alpha,j} \frac{A_{0}}{N_{\perp}} {\bf S}_{\alpha} (r_{j}) \cdot {\bf \tilde{I}}_{\alpha} (r_{j}) 
\label{Eq:H_hf}
\end{equation}
where $\alpha=A,B$ denotes the sublattice index, $j=1, \ldots, N$ is the site index of cross sections along the tube axis, and
$A_{0}$ is the hyperfine coupling constant. There is a discrepancy between the measured hyperfine coupling constant and the theoretical prediction. The observed value in an isotopically enriched ($\sim 99\%$ $^{13}$C) nanotube quantum dots~\cite{Churchill:2009a,Churchill:2009b} was two orders larger than the theoretical calculation employing a noninteracting system calculation.~\cite{Fischer:2009,Yazyev:2008,Csiszar:2014}
Whereas the measured value was extracted through theories developed for other materials without valley degrees of freedom,~\cite{Jouravlev:2006} such as GaAs, and needs to be further confirmed,~\cite{Trauzettel:2009} we take $A_{0}=6.0~\mu$eV, which is in the order between the observed and theoretical values, for the purpose of estimation. We also note that subsequent measurements in CNTs with natural abundance ($\sim 1\%~^{13}$C)~\cite{Pei:2012,Laird:2013} corroborate the hyperfine coupling constant reported in Refs.~\citenum{Churchill:2009a,Churchill:2009b}. 

We split the nanotube into small cylinders of height $a$, the length of the atomic scale, and get $N_{\perp}=\pi R a n_{I}$ as the number of the atoms on each sublattice in such a cylinder, with $n_I$ the atomic area density of a graphene sheet. We group the nuclear spins within one cylinder into an effective composite spin, ${\bf \tilde{I}}_{\alpha} (r_{j}) \equiv \sum_{j_{\perp}} {\bf I}_{\alpha}({\bf r}_{j,j_{\perp}})$, which we refer to as a spin of a cross section. Because $N_{\perp}\gg 1$, the effective spins are large, with maximal magnitude $N_{\perp}I$, and thus can be treated semiclassically. 
We choose $a$ to be the lattice constant of the CNT (the carbon-carbon bond length is $a/\sqrt{3}$), which for $(6,6)$ CNTs gives $N_{\perp} = 12$ and the radius $R\approx 4.1$~{\AA}. The choice of $(6,6)$ CNTs is partially motivated by the experiment in Ref.~\citenum{Sanchez-Valencia:2014}, which reports that defect-free CNTs with definite chiral index have been made possible.
We also denote the nuclear spin of $^{13}$C as $I=\frac{1}{2}$. Further, the effective one-dimensional electron spin density operator is 
\begin{eqnarray}
{\bf S}_{\alpha} (r) = \frac{N_{\perp}}{n_{I}} \left| \psi_{\perp,\alpha} \right|^2 \sum_i \delta(r-r_{i}) ~ \boldsymbol{\sigma}_i,
\end{eqnarray}
where $r$ denotes the coordinate along the tube, $r_i$ the position operator of the $i$th electron, $\boldsymbol{\sigma}_i $ is a vector with components formed by the Pauli matrices in spin space of the $i$th electron, and $\psi_{\perp,\alpha}$ is the transverse part of the electron wave function (assumed to be the same for all
electrons). We will assume $\psi_{\perp,\alpha}$ spreads uniformly over the circumference, so that $\left| \psi_{\perp,\alpha} \right|^2=1/(2\pi R)$. 

Since $A_{0}\ll\epsilon_{F}$, we use the Schrieffer-Wolff transformation to integrate out the electron degrees of freedom,~\cite{Schrieffer:1966,Simon:2007,Simon:2008} which results in an effective RKKY interaction between two localized spins,
\begin{eqnarray}
\mathcal{H}_{\textrm{RKKY}} &=& \frac{1}{N_{\perp}^2} \sum_{i,j,\alpha,\beta} \sum_{\mu\nu} J_{\alpha\beta}^{\mu\nu}(r_i-r_j) \tilde{I}_{\alpha}^{\mu} (r_{i})\tilde{I}_{\beta}^{\nu} (r_{j}),
\label{Eq:RKKY_real}
\end{eqnarray}
where $\mu,\nu=x,y,z$ are coordinates in spin space and the effective RKKY exchange coupling,
\begin{equation}
J_{\alpha\beta}^{\mu\nu}(r_i-r_j) = \frac{A_0^2a^2}{2} \chi_{\alpha\beta}^{\mu\nu}(r_i-r_j).
\label{Eq:J_chi_real}
\end{equation} 
The static spin susceptibility is defined as
\begin{equation}
\chi_{\alpha\beta}^{\mu\nu}(r_i-r_j)=-\frac{i}{a^2}\int_{0}^{\infty} dt \; e^{-\eta t} \left\langle \left[S_{\alpha}^{\mu}(r_i,t),S_{\beta}^{\nu}(r_j,0)\right] \right\rangle,
\label{Eq:susceptibility}
\end{equation}
with an infinitesimal positive $\eta$ and $\langle ... \rangle$ being the average corresponding to the one-dimensional effective electron Hamiltonian $\mathcal{H}_{\textrm{el}}$. For the continuum description, we will replace $S_{\alpha}^{\mu}(r_i,t)/a$ with the operators $S_{\alpha}^{\mu}(r,t)$.

\subsection{Electron Hamiltonian and bosonization \label{SubSec:Hel}}

In this section, we discuss the one-dimensional effective electron Hamiltonian and its bosonized form. We start with the Hamiltonian of an interacting electronic system,
\begin{equation}
\mathcal{H}_{\textrm{el}}=\mathcal{H}_{0}+\mathcal{H}_{\textrm{int}},
\end{equation}
where $\mathcal{H}_{0}$ and $\mathcal{H}_{\textrm{int}}$ describe the kinetic energy and interaction terms, respectively. 

The Hamiltonian $\mathcal{H}_{0}$ is defined by a tight-binding model of a carbon lattice, including the nearest-neighbor hopping terms with the hopping parameter $t$. We neglect the longer-range hopping, nanotube curvature, and spin-orbit interactions, which results in a Hamiltonian conserving the total spin. The spin susceptibility can then be written as
$\chi_{\alpha\beta}^{\mu\nu}(r_{i}-r_{j})=\delta_{\mu\nu}\chi_{\alpha\beta}^{\mu}(r_{i}-r_{j})$.~\cite{Klinovaja:2013a} 
We Fourier transform $\mathcal{H}_{0}$ and expand it around the Dirac points, ${\bf K}_{\gamma} = \gamma k_{v} \hat{z} + \frac{2\pi}{\sqrt{3}a} \hat{t}$ with $k_{v} \equiv \frac{2\pi}{3 a}$ ($\hat{z}$ and $\hat{t}$ being the unit vectors along the tube axis and circumference, respectively).~\footnote[3]{Our definition for $k_{v}$ differs from the one in Refs.~\citenum{Braunecker:2009a,Braunecker:2009b} by a reciprocal-lattice vector.} 
With the assumption that the conduction electrons are confined into the lowest transverse mode due to the spatial confinement,
the tight-binding model results in~\cite{Saito:1992} 
\begin{eqnarray}
\mathcal{H}_{0} &=& \sum_{q,\gamma,\sigma} \left(
c_{A,\gamma,\sigma}^{\dagger}(q) \; c_{B,\gamma,\sigma}^{\dagger}(q)
 \right) \nonumber\\
&& \hspace{0.2in} \times 
\left(
\begin{array}{cc}
0 &  -\gamma \hbar v_{F} q \\
-\gamma \hbar v_{F} q & 0
\end{array}
\right) \left(
\begin{array}{c}
c_{A,\gamma,\sigma}(q) \\
c_{B,\gamma,\sigma}(q)
\end{array}
\right),
\label{Eq:H0}
\end{eqnarray}
where $c_{\alpha,\gamma,\sigma}^{\dagger}(q)$ is the creation operator with the sublattice index $\alpha=A,B$ ($\alpha=\pm 1$), valley index $\gamma=\pm$, spin $\sigma=\uparrow,\downarrow$, the $z$ component of the momentum $q=q_{z}$ is measured from the Dirac point ${\bf K}_{\gamma}$, and $v_{F}=\frac{\sqrt{3}ta}{2\hbar}$ is the Fermi velocity. 
Eq.~(\ref{Eq:H0}) can be diagonalized by symmetric ($\delta=+$) and antisymmetric ($\delta=-$) combinations,
\begin{equation}
\psi_{\delta,\gamma,\sigma} (q) = \frac{1}{\sqrt{2}} \left[
c_{A,\gamma,\sigma} (q) + \delta c_{B,\gamma,\sigma} (q) \right],
\label{Eq:diag}
\end{equation}
corresponding to the eigenvalues, $E_{\delta\gamma}=-\delta \gamma  \hbar v_{F} q$. Therefore, the energy spectrum of Eq.~(\ref{Eq:H0}) exhibits linear dispersions close to the Dirac points, leading to two copies of Luttinger liquid spectrum located at $k_{z}=\pm k_{v}$ (see Fig.~\ref{Fig:scattering}).

To proceed, we describe the system in terms of the right ($R\equiv +1$) and left ($L\equiv -1$) moving particles,
$\psi_{\ell,\gamma,\sigma} (q)$, where for $\ell=R$ and $L$, we have 
\begin{subequations}
\label{Eq:LRmovers}
\begin{eqnarray}
\psi_{R,\gamma,\sigma} (q) &\equiv& \left.\psi_{\delta,\gamma,\sigma} (q)\right|_{\delta=-\gamma}, \\
\psi_{L,\gamma,\sigma} (q) &\equiv& \left.\psi_{\delta,\gamma,\sigma} (q)\right|_{\delta=\gamma}, 
\end{eqnarray}
\end{subequations}
respectively. Combining Eqs.~(\ref{Eq:diag}) and (\ref{Eq:LRmovers}), we find the relation between the original electron operators and the right/left movers, 
\begin{equation}
c_{\alpha,\gamma,\sigma} (q) = \frac{1}{\sqrt{2}} \left[ \left.\psi_{\ell,\gamma,\sigma} (q)\right|_{\ell=-\gamma} + \alpha \left.\psi_{\ell,\gamma,\sigma} (q)\right|_{\ell=\gamma} \right].
\label{Eq:transform}
\end{equation}
One may bosonize $\psi_{\ell,\gamma,\sigma}$ in terms of the bosonic fields $\theta_{\gamma\sigma}$ and $\phi_{\gamma\sigma}$.~\cite{Giamarchi:2003,Bruus:2004} In real space, we have 
\begin{equation}
\psi_{\ell,\gamma,\sigma} (r)= \frac{U_{\ell,\gamma,\sigma}}{\sqrt{2\pi a}} e^{i (\ell k_{F}+\gamma k_{v})r} e^{-i \left[ \ell \phi_{\gamma\sigma}(r) - \theta_{\gamma\sigma}(r) \right]},
\label{Eq:bosonization}
\end{equation}
where $k_{F}$ is the Fermi wave number, the lattice constant $a$ sets the smallest length scale of the system, $U_{\ell,\gamma,\sigma}$ is the Klein factor removing a $(\ell,\gamma,\sigma)$ particle from the system, and the bosonic fields satisfy the following relations~\cite{Giamarchi:2003}
\begin{eqnarray}
\left[ \phi_{\gamma\sigma}(r_{1}), \theta_{\gamma'\sigma'}(r_{2})\right] &=& i \frac{\pi}{2} \textrm{sign}(r_{2}-r_{1}) \delta_{\gamma\gamma'}\delta_{\sigma\sigma'}, \\
\triangledown \phi_{\gamma\sigma}(r) &=& -\pi \left[ \rho_{R,\gamma,\sigma} (r) + \rho_{L,\gamma,\sigma} (r) \right], \\
\triangledown \theta_{\gamma\sigma}(r) &=& \pi \left[ \rho_{R,\gamma,\sigma} (r) - \rho_{L,\gamma,\sigma} (r) \right],
\end{eqnarray}
with the real-space density operator $\rho_{\ell, \gamma,\sigma}(r) = \psi_{\ell,\gamma,\sigma}^{\dagger}(r) \psi_{\ell,\gamma,\sigma}(r)$. One can see that the field $\triangledown \theta_{\gamma\sigma}(r) / \pi$ is canonically conjugate to $\phi_{\gamma\sigma}(r)$,
\begin{equation}
\left[ \phi_{\gamma\sigma}(r_{1}), \frac{\triangledown \theta_{\gamma'\sigma'}(r_{2})}{\pi} \right] = i \delta(r_{2}-r_{1}) \delta_{\gamma\gamma'}\delta_{\sigma\sigma'}.
\end{equation}

Including the electron-electron interaction, the electron Hamiltonian $\mathcal{H}_{\textrm{el}}$ can then be bosonized~\cite{Kane:1997,Egger:1997,Egger:1998} 
\begin{eqnarray}
\mathcal{H}_{\textrm{el}} &=& \sum_{\nu,P} \int \frac{\hbar dr}{2\pi} \left\{
u_{\nu P} K_{\nu P} \left[ \triangledown \theta_{\nu P}(r) \right]^2
 \right. \nonumber \\
&& \hspace{1.2in} \left. 
+ \frac{u_{\nu P}}{K_{\nu P}} \left[ \triangledown \phi_{\nu P}(r) \right]^2
\right\},
\label{Eq:LL}
\end{eqnarray}
where $\nu \in \{c\equiv +,s\equiv-\}$ refers to the charge/spin sectors, and $P \in \{S\equiv+,A\equiv-\}$ the symmetric/antisymmetric combination of the bosonic fields between the $\gamma=\pm$ valleys, namely,
\begin{eqnarray}
\theta_{\nu P} &\equiv& 
\frac{1}{2} \left[\theta_{+,\uparrow} + \nu \theta_{+,\downarrow}
+ P \left( \theta_{-,\uparrow} + \nu \theta_{-,\downarrow} \right)\right], \label{Eq:theta} \\
 \phi_{\nu P} &\equiv& \frac{1}{2} \left[
\phi_{+,\uparrow} + \nu \phi_{+,\downarrow} + P \left( \phi_{-,\uparrow} + \nu \phi_{-,\downarrow} \right)
\right]. \label{Eq:phi}
\end{eqnarray}
The velocities for the $(\nu,P)$ channels are $u_{\nu P}=v_{F}/K_{\nu P}$. 
The indices defined in this section are summarized in Table~\ref{Tab:indices} for reference.
\begin{table}[t]
\begin{tabular}[c]{ | c | c | c | }
\hline
~Index~ & Degree of freedom & Possible values  \\
\hline
$\alpha$ & sublattice & ~$A$ ($\equiv +1$), $B$ ($\equiv -1$)~  \\ 
\hline
$\gamma$ & valley & $+$, $-$  \\
\hline
$\sigma$ & spin  & $\uparrow$ ($\equiv +1$), $\downarrow$ ($\equiv -1$)  \\
\hline
$\delta$ & \begin{tabular}{@{}c@{}}~symmetric/antisymmetric~ \\ combination of $\alpha=A,B$ \end{tabular} & $+$, $-$  \\
\hline
$\ell$ & right/left mover & $R$ ($\equiv +1$), $L$ ($\equiv -1$)  \\
\hline
$\nu$ & charge/spin sector & $c$ ($\equiv +1$), $s$ ($\equiv -1$)  \\
\hline
$P$ & \begin{tabular}{@{}c@{}}symmetric/antisymmetric \\ combination of $\gamma=\pm$ \end{tabular} & $S$ ($\equiv +1$), $A$ ($\equiv -1$)  \\
\hline
\end{tabular}
\caption{
The indices defined in Sec.~\ref{Sec:RKKY}, the corresponding degrees of freedom, and the possible values of the indices.}
\label{Tab:indices}
\end{table}
The noninteracting case corresponds to the Luttinger liquid parameters $K_{cS}=K_{cA}=K_{sS}=K_{sA}=1$, and the repulsive electron-electron interaction leads to $K_{cS}<1$. 
The parameter $K_{cS}$ depends on the radius of CNTs through the relation, $K_{cS}=\left[1+ (8e^2)/(\pi \hbar v_{F}) \ln \left( R_{s}/R \right) \right]^{-\frac{1}{2}}$, where $e$ is the electron charge and $R_{s}\approx 1000~$\AA~ is the screening length.~\cite{Kane:1997} Therefore, for CNTs with smaller radius, the electron-electron interaction has stronger effects due to the stronger spatial confinement, as expected. However, this radius dependence is relatively weak because of its logarithmic form. For $R=4.1$-$100~$\AA, $K_{cS}\approx 0.16$-$0.24$. In this work, we take
$K_{cS} \approx 0.2$ and $K_{cA}\approx K_{sS} \approx K_{sA} \approx 1$.~\cite{Kane:1997,Egger:1998,Bockrath:1999,Ishii:2003} 
 
With Eqs.~(\ref{Eq:bosonization}), (\ref{Eq:theta}), and (\ref{Eq:phi}), we can write the single-particle spin operator $S_{\alpha}^{\mu}(r)$ in terms of the bosonic fields to compute the correlation functions in Eq.~(\ref{Eq:susceptibility}). Since the electron Hamiltonian (\ref{Eq:LL}) is a free bosonic system, the correlation functions can be computed straightforwardly within the Luttinger liquid formalism.~\cite{Giamarchi:2003}

\subsection{Spin operator in terms of the bosonic fields \label{SubSec:spin}}

To examine the sublattice dependence, we first write the spin operator in terms of the original electron operators with the explicit sublattice index $\alpha$. 
\begin{eqnarray}
S_{\alpha}^{\mu}(r_{j}) &\equiv& \frac{1}{2} \sum_{\sigma,\sigma'} \sum_{\gamma,\gamma'} \sigma_{\sigma\sigma'}^{\mu} c_{\alpha,\gamma,\sigma}^{\dagger}(r_{j}) c_{\alpha,\gamma',\sigma'}(r_{j}),
\end{eqnarray}
which, according to Eq.~(\ref{Eq:transform}), can be written as $S_{\alpha}^{\mu}(r_{j}) = S_{f,\alpha}^{\mu}(r_{j}) + S_{b,\alpha}^{\mu}(r_{j})$, where 
\begin{eqnarray}
S_{f,\alpha}^{\mu}(r_{j}) &\equiv& \frac{1}{4} \sum_{\sigma,\sigma'} \sigma_{\sigma\sigma'}^{\mu} \sum_{\ell,\gamma} \left[ \psi_{\ell,\gamma,\sigma}^{\dagger}(r_{j}) \psi_{\ell,\gamma,\sigma'}(r_{j})
\right. \nonumber\\
&& \hspace{0.75in} \left.
+ \alpha \psi_{\ell,\gamma,\sigma}^{\dagger}(r_{j}) \psi_{\ell,\bar{\gamma},\sigma'}(r_{j}) \right], 
\end{eqnarray}
arises from the forward scattering ($q \sim 0$ or $q \sim 2k_{v}$) and
\begin{eqnarray}
S_{b,\alpha}^{\mu}(r_{j}) &\equiv& \frac{1}{4} \sum_{\sigma,\sigma'} \sigma_{\sigma\sigma'}^{\mu} \sum_{\ell,\gamma} \left[ \alpha \psi_{\ell,\gamma,\sigma}^{\dagger}(r_{j}) \psi_{\bar{\ell},\gamma,\sigma'}(r_{j})
\right. \nonumber\\
&& \hspace{0.85in} \left.
 + \psi_{\ell,\gamma,\sigma}^{\dagger}(r_{j}) \psi_{\bar{\ell},\bar{\gamma},\sigma'}(r_{j}) \right],
\label{Eq:spin_op_back}
\end{eqnarray}
corresponds to the back scattering [$q \sim 2k_{F}$ or $q \sim 2(k_{v} \pm k_{F})$]. Here $\sigma_{\sigma\sigma'}^{\mu}$ are the Pauli matrices in spin space, $\bar{\ell} \equiv -\ell$, $\bar{\gamma} \equiv -\gamma$, and the inverse Fourier transform of $\psi_{\ell,\gamma,\sigma}(q)$ is given by
\begin{equation}
\psi_{\ell,\gamma,\sigma}(r_{j})=\frac{1}{\sqrt{N}} \sum_{q} e^{iqr_{j}}\psi_{\ell,\gamma,\sigma}(q),
\end{equation}
with $\psi_{\ell,\gamma,\sigma}(q)$ defined in Eq.~(\ref{Eq:LRmovers}). 

\begin{figure}[htb]
\centering
\includegraphics[width=\linewidth]{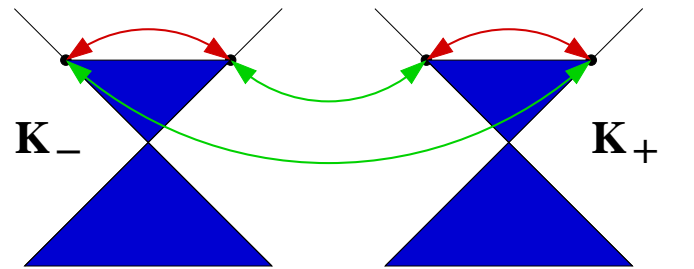} 
\caption{Back scattering processes on the Fermi surface. $K_{\gamma}$ indicates the two Dirac points with the valley index $\gamma=\pm$. 
The red (green) arrows correspond to the intravalley back scattering $\psi_{\ell,\gamma,\sigma}^{\dagger} \psi_{\bar{\ell},\gamma,\sigma'}$ (intervalley back scattering $\psi_{\ell,\gamma,\sigma}^{\dagger} \psi_{\bar{\ell},\bar{\gamma},\sigma'}$) processes. The arrows are mutually independent and the spins are not shown.}
\label{Fig:scattering}
\end{figure}

Since we consider the temperature $T$ much lower than the Fermi energy $\epsilon_{F}$, the states below $\epsilon_{F}$ are filled, allowing us to keep only the scattering processes that take place on the Fermi surface. 
In contrast to the back scattering term, the scaling dimension of the forward scattering term does not depend on $K_{cS}$, which is the only Luttinger parameter modified by the electron-electron interaction, so the forward scattering term produces only local extrema (peaks) in the RKKY interaction.~\cite{Giamarchi:2003,Braunecker:2009a,Braunecker:2009b} 
Since the nuclear spin order is determined by the global extrema of the RKKY interaction, in what follows we may neglect $S_{f,\alpha}^{\mu}(r_{j})$, and focus on the back scattering term, $S_{b,\alpha}^{\mu}(r_{j})$.
Each term of Eq.~(\ref{Eq:spin_op_back}) corresponds to a scattering process, as illustrated in Fig.~\ref{Fig:scattering}. 

Taking the continuum limit, we obtain
\begin{eqnarray}
S_{b,\alpha}^{\mu}(r) &=& 
S_{b,\mathrm{intra},\alpha}^{\mu}(r) +S_{b,\mathrm{inter},\alpha}^{\mu} (r), \label{Eq:spin_operator} \\
S_{b,\mathrm{intra},\alpha}^{\mu} (r) &\equiv&
\frac{\alpha}{4} \sum_{\sigma,\sigma'} \sigma_{\sigma\sigma'}^{\mu} 
\sum_{\gamma,\ell}
\left[ 
\psi_{\ell,\gamma,\sigma}^{\dagger} (r) \psi_{\bar{\ell},\gamma,\sigma'} (r)
\right], \label{Eq:spin_operator_intra} \\
S_{b,\mathrm{inter},\alpha}^{\mu} (r) &\equiv& 
\frac{1}{4} \sum_{\sigma,\sigma'} \sigma_{\sigma\sigma'}^{\mu} 
\sum_{\gamma,\ell}
\left[
\psi_{\ell,\gamma,\sigma}^{\dagger} (r) \psi_{\bar{\ell},\bar{\gamma},\sigma'} (r)
\right]. \label{Eq:spin_operator_inter}
\end{eqnarray}
From Eq.~(\ref{Eq:spin_operator_intra}) we see that the intravalley back scattering term, $S_{b,\mathrm{intra},\alpha}^{\mu}$, is opposite for the two sublattices. 
We will see below that this term gives rise to the $q=2 k_{F}$ RKKY peak in the spin susceptibility. The intervalley back scattering terms in $S_{b,\mathrm{inter},\alpha}^{\mu}$, on the other hand, do not depend on the sublattice index and give rise to $q=2(k_{F} - \gamma k_{v})$ peaks. 
The spin susceptibility in the absence of the feedback thus contains two parts, the sublattice-dependent $q=2k_{F}$ intravalley back scattering, and $q=2(k_{F} - \gamma k_{v})$ intervalley back scattering, which is independent of sublattice, as in Refs.~\citenum{Saremi:2007,Brey:2007,Black-Schaffer:2010,Kogan:2011,Klinovaja:2013a,Power:2013}. However, the latter will be suppressed when taking into account the feedback (Overhauser field due to ordered nuclear spins), as will be discussed in Sec.~\ref{Sec:feedback}. As a result, in spite of its presence in the spin susceptibility, the intervalley back scattering will not influence the nuclear spin order established by the intravalley back scattering. 
For clarity, we list the scattering processes in $S^{\mu}_{\alpha}$, their operators, and the locations and types of the corresponding extrema in the RKKY interaction in Table~\ref{Tab:scattering}.
\begin{table}[htb]
\begin{tabular}[c]{ | c | c | c | c | }
\hline
Scattering process & Operator & Location & Type \\
\hline
~Intravalley, forward~ & ~$\psi_{\ell,\gamma,\sigma}^{\dagger} \psi_{\ell,\gamma,\sigma'}$~ & $q \sim 0$ & local \\ [0.02in]
\hline
Intervalley, forward & $\psi_{\ell,\gamma,\sigma}^{\dagger} \psi_{\ell,\bar{\gamma},\sigma'}$ & $q \sim 2k_{v}$ & local \\ [0.02in]
\hline
Intravalley, back & $\psi_{\ell,\gamma,\sigma}^{\dagger} \psi_{\bar{\ell},\gamma,\sigma'}$ & $q \sim 2k_{F}$ & ~global~ \\ [0.02in]
\hline
Intervalley, back & $\psi_{\ell,\gamma,\sigma}^{\dagger} \psi_{\bar{\ell},\bar{\gamma},\sigma'}$ & ~$q \sim 2(k_{v} \pm k_{F})$~ & local \\ [0.02in]
\hline
\end{tabular}
\caption{
The scattering processes in $S^{\mu}_{\alpha}$, their corresponding operators, and the locations and types of the corresponding extrema in the RKKY interaction in the absence of the feedback.}
\label{Tab:scattering}
\end{table}

From now on we shall proceed with the intravalley back scattering term, $S_{b,\textrm{intra},\alpha}^{\mu}$, and will come back to the intervalley back scattering term when discussing the feedback in Sec.~\ref{SubSec:Overhauser}.
Now the spin operator is expressed in terms of the operators $\psi_{\ell,\gamma,\sigma} (r)$, which can be bosonized through Eqs.~(\ref{Eq:bosonization}), (\ref{Eq:theta}), and (\ref{Eq:phi}).
To this end, we define the spin density wave operators,~\cite{Giamarchi:2003}
\begin{equation}
O_{SDW,\gamma}^{\mu}(r) \equiv \sum_{\sigma,\sigma'} \sigma_{\sigma\sigma'}^{\mu}  
\psi_{R,\gamma,\sigma}^{\dagger} (r) \psi_{L,\gamma,\sigma'} (r),
\end{equation}
such that
\begin{equation}
S_{b,\mathrm{intra},\alpha}^{\mu} (r) = \frac{\alpha}{4} \sum_{\gamma} \left[ O_{SDW,\gamma}^{\mu}(r) + \left( O_{SDW,\gamma}^{\mu}(r) \right)^{\dagger} \right],
\label{Eq:SDW_operator}
\end{equation}
which can be written in terms of the bosonic operators, $\theta_{\nu P}$ and $\phi_{\nu P}$,
\begin{widetext}
\begin{eqnarray}
\left[ O_{SDW,\gamma}^{x}({r}) \right]^{\dagger}
&=&  \frac{1}{2\pi a} e^{2ik_{F}r} \left\{
 e^{-i \left[ \phi_{cS}({r}) +\gamma \phi_{cA}({r}) + \theta_{sS} ({r}) + \gamma \theta_{sA}({r}) \right]}
+ e^{-i \left[ \phi_{cS}({r}) +\gamma \phi_{cA}({r}) - \theta_{sS} ({r}) - \gamma \theta_{sA}({r}) \right]} \right\}, \label{Eq:SDW_x} \\
\left[ O_{SDW,\gamma}^{y}({r}) \right]^{\dagger}
&=& \frac{-i}{2\pi a} e^{2ik_{F}r} \left\{
 e^{-i \left[ \phi_{cS}({r}) +\gamma \phi_{cA}({r}) + \theta_{sS} ({r}) + \gamma \theta_{sA}({r}) \right]}
- e^{-i \left[ \phi_{cS}({r}) +\gamma \phi_{cA}({r}) - \theta_{sS} ({r}) - \gamma \theta_{sA}({r}) \right]} \right\}, \label{Eq:SDW_y} \\
\left[ O_{SDW,\gamma}^{z}({r}) \right]^{\dagger}
&=& \frac{1}{2\pi a} e^{2ik_{F}r} \left\{
 e^{-i \left[ \phi_{cS}({r}) +\gamma \phi_{cA}({r}) + \phi_{sS} ({r}) + \gamma \phi_{sA}({r}) \right]}
- e^{-i \left[ \phi_{cS}({r}) +\gamma \phi_{cA}({r}) - \phi_{sS} ({r}) - \gamma \phi_{sA}({r}) \right]} \right\},  \label{Eq:SDW_z}
\end{eqnarray}
\end{widetext}
where the Klein factors $U_{\ell,\gamma,\sigma}$ are omitted because they have no influence.

\subsection{Spin susceptibility and RKKY interaction\label{SubSec:chi}}

With Eqs.~(\ref{Eq:SDW_operator})--(\ref{Eq:SDW_z}), the spin susceptibility can be expressed in terms of the bosonic fields, and calculated within the Luttinger liquid formalism. First, let 
\begin{equation}
\chi_{\alpha\beta}^{>,\mu}(r,t) \equiv -i \left\langle S_{b,\mathrm{intra},\alpha}^{\mu}(r,t) S_{b,\mathrm{intra},\beta}^{\mu}(0) \right\rangle,
\label{Eq:chi_greater}
\end{equation}
where $\langle \cdots \rangle$ denotes the time-ordered average corresponding to the electron Hamiltonian, Eq.~(\ref{Eq:LL}), and the time argument appears due to the interaction representation adopted for the operators. In the continuum limit we have~\cite{Giamarchi:2003}
\begin{eqnarray}
\chi_{\alpha\beta}^{\mu}(r) &=& -2i \int_{0}^{\infty} dt \; e^{-\eta t} \; \Theta(t) \; \textrm{Im} \left[
i \chi_{\alpha\beta}^{>,\mu}(r,t)\right].
\end{eqnarray}
Since Eq.~(\ref{Eq:LL}) is a free bosonic Hamiltonian, the calculation of the correlation functions is rather straightforward.~\cite{Giamarchi:2003} 
Upon the Fourier transform, $\chi_{\alpha\beta}^{\mu}(q) = \int dr e^{-iqr} \chi_{\alpha\beta}^{\mu}(r)$,~\footnote[4]{Here we adopt the definition of $\chi^{\mu}(q)$ in Ref.~\citenum{Braunecker:2009b}, where $\chi^{\mu}(q)$ has a different dimension than $\chi^{\mu}(r)$, whereas both $J^{\mu}(q)$ and $J^{\mu}(r)$ have the dimension of energy. Therefore, the form of Eq.~\eqref{Eq:J_chi_real} is different from Eq.~\eqref{Eq:JRKKY} by a factor of $a$.}   
the static spin susceptibility in momentum space reads
\begin{eqnarray}
\chi_{AA}^{\mu}(q) &=& -\chi_{AB}^{\mu}(q) \nonumber\\
&=& - \frac{ \sin (\pi g_{\mu}) }{(4\pi)^2 \hbar v_{F}}
\left(\frac{\lambda_{T}}{2\pi a} \right)^{2-2g_{\mu}} \nonumber\\
&& \times 
\sum_{\kappa=\pm}
\left| 
\frac{ \Gamma\left( 1-g_{\mu}\right) \Gamma\left( \frac{g_{\mu}}{2}-i\frac{\lambda_{T}}{4\pi}\left( 
q - 2 \kappa k_{F} \right) \right)}
{\Gamma\left(
\frac{2-g_{\mu}}{2}-i\frac{\lambda_{T}}{4\pi}\left( q - 2 \kappa k_{F} \right) \right)}
\right|^{2}, \nonumber\\
\label{Eq:chi}
\end{eqnarray}
where we have defined the thermal length $\lambda_{T}=\frac{\hbar v_{F}}{k_{B}T}$, and $\Gamma(x)$ is the Gamma function. The spin susceptibility strongly depends on the exponents,
\begin{eqnarray}
g_{x}=g_{y} &=& \frac{1}{4}\left( K_{cS} + K_{cA} + \frac{1}{K_{sS}} + \frac{1}{K_{sA}} \right), \\
g_{z} &=&\frac{1}{4}\left( K_{cS} + K_{cA} + K_{sS} + K_{sA} \right).
\end{eqnarray}
For the systems with the spin rotational symmetry, we have $K_{sS}=K_{sA}=1$, which leads to $g_{x}=g_{y}=g_{z}$ and thus isotropic spin susceptibility, as expected. 

In Eq.~\eqref{Eq:chi} we obtained the {\it opposite} sign for $\chi_{AB}^{\mu}(q)$ explicitly. The antiferromagnetic correlation between spins on different sublattice sites provides a consistent picture with the noninteracting calculation.~\cite{Saremi:2007,Brey:2007,Black-Schaffer:2010,Kogan:2011,Klinovaja:2013a,Power:2013} 
Our results thus consistently reconcile Refs.~\citenum{Braunecker:2009a,Braunecker:2009b} and \citenum{Saremi:2007,Brey:2007,Black-Schaffer:2010,Kogan:2011,Klinovaja:2013a,Power:2013}, which obtained the interaction-induced boost for the transition temperature, and the locally antiferromagnetic coupling, separately.

\begin{figure}[htb]
\centering
\includegraphics[width=\linewidth]{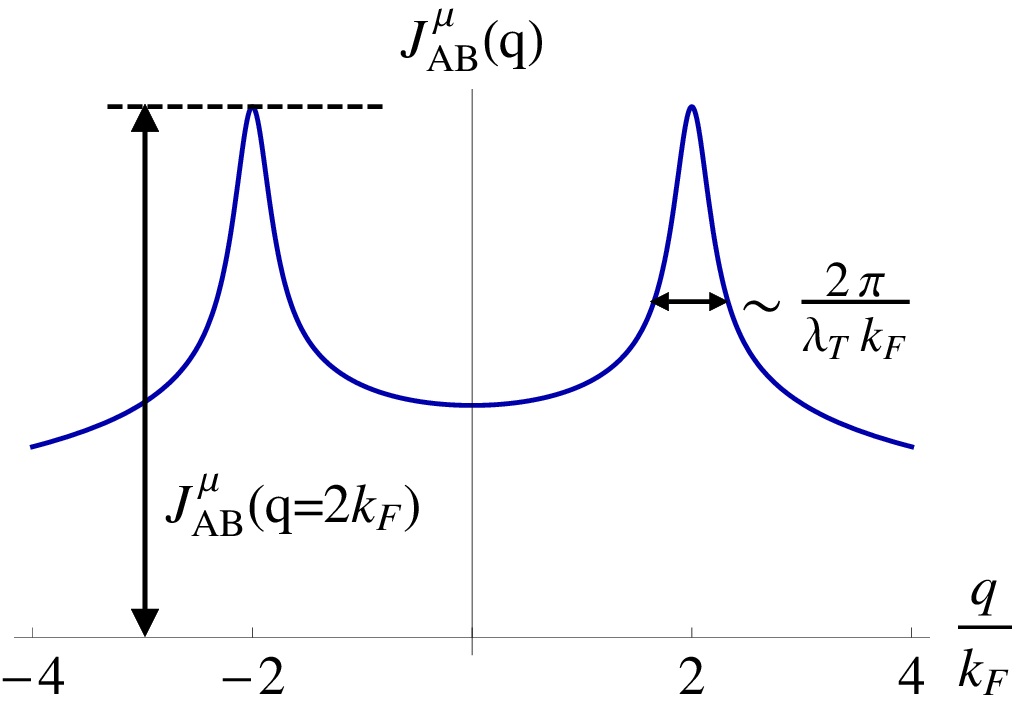}
\caption{RKKY interaction from Eq.~(\ref{Eq:JRKKY}) in momentum space. The interaction has peaks at $q=\pm 2k_{F}$ with the width $\sim\frac{2\pi}{\lambda_{T}k_{F}}$. The parameters used here are $K_{cS} = 0.2$, $K_{cA}=K_{sS}=K_{sA}=1$,~\cite{Pennington:1996,Kane:1997,Egger:1997,Egger:1998,Bockrath:1999,Ishii:2003} $I=\frac{1}{2}$, $A_{0}= 6.0~\mu$eV,~\cite{Yazyev:2008,Churchill:2009a,Churchill:2009b,Fischer:2009,Pei:2012,Laird:2013,Csiszar:2014} $v_{F}= 8.0\times 10^5~$m/s, $k_{F}= 4.0\times 10^8~$m$^{-1}$, $a= 2.46~${\AA},~\cite{Braunecker:2009b} $L= 1.0~\mu$m, and $N_{\perp}= 12$.~\cite{Sanchez-Valencia:2014} 
For the purpose of illustration, we choose an unrealistically short thermal length, corresponding to an unrealistically high temperature $T=100~$K, to demonstrate the RKKY peaks. For realistic temperatures, the peaks will be much sharper.}
\label{Fig:RKKY}
\end{figure}

Since the RKKY coupling is related to the spin susceptibility by 
\begin{equation}
J_{\alpha \beta}^\mu(q) = \frac{A_0^2 a}{2} \chi_{\alpha \beta}^\mu(q),
\label{Eq:JRKKY}
\end{equation}
we can use Eq.~\eqref{Eq:chi} to evaluate the RKKY coupling. Its momentum dependence is plotted in Fig.~\ref{Fig:RKKY}. 
We remind that Eq.~\eqref{Eq:chi} contains only the contribution from the $q \sim 2k_{F}$ intravalley back scattering terms, leading to the global extrema. The contributions from other scattering processes only give local extrema, and will be suppressed in the presence of the feedback (see Table~\ref{Tab:scattering}). 
The peak value of the RKKY interaction is given by
\begin{eqnarray}
&& J_{AB}^\mu(q=2k_{F}) \nonumber\\
\approx &&  
\frac{ A_0^2 a \sin (\pi g_{\mu}) }{32\pi^2 \hbar v_{F}}
\left( \frac{\lambda_{T}}{2\pi a} \right)^{2-2g_{\mu}}
\left| 
\frac{ \Gamma\left( 1-g_{\mu}\right) \Gamma\left( \frac{g_{\mu}}{2} \right)}
{\Gamma\left(\frac{2-g_{\mu}}{2} \right)} \right|^{2}, 
\label{Eq:peak}
\end{eqnarray}
which depends on the temperature through the thermal length, $\lambda_{T}$.

\section{Antiferromagnetic nuclear spin helix \label{Sec:helical}}
\subsection{Antiferromagnetic helix and magnon spectrum \label{SubSec:magnon}}

We now perform the spin-wave analysis to find spectrum of the low-energy excitations of the RKKY Hamiltonian in Eq.~\eqref{Eq:RKKY_real}.~\cite{Holstein:1940,VanKranendonk:1958} Since the RKKY interaction in CNTs is sublattice dependent, it leads to a different nuclear spin order from the ferromagnetic helical order in GaAs nanowires. We first consider only the long-wavelength magnons propagating along the tube axis, and will include short-wavelength magnon excitations when estimating the transition temperature in Sec.~\ref{SubSec:Tc}. 

We begin by assuming that in a given cross section (i.e. along the transverse direction) the nuclear spins on the same sublattice sites in the ground state are parallel to each other and the spins on different sublattice sites point to the opposite direction, $\tilde{{\bf I}}_{A}(r_{j})=-\tilde{{\bf I}}_{B}(r_{j})$.
A helical order means these spins rotate within a fixed plane as one moves along the tube, with a spatial period $\pi/k_{F}$. We denote this (helical) plane as $xy$ and the axis perpendicular to it as $z$.
The confinement of the nuclear spins to the $xy$ plane will be justified in Sec.~\ref{Sec:feedback}, where we will show the modified RKKY interaction due to the feedback to be anisotropic: $|\tilde{J}_{\alpha\beta}^{x}(q)|=|\tilde{J}_{\alpha\beta}^{y}(q)|>|\tilde{J}_{\alpha\beta}^{z}(q)|$.

\begin{figure}[htb]
\centering
\includegraphics[width=\linewidth]{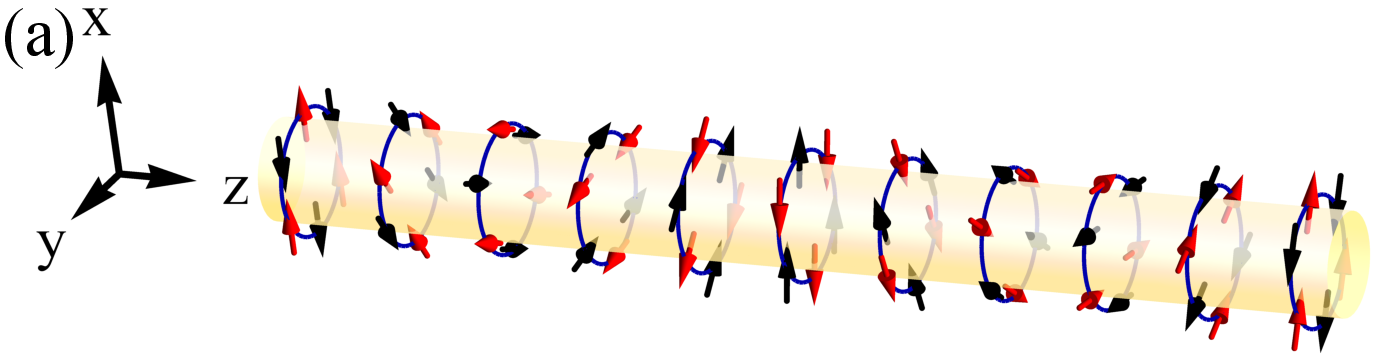}
\includegraphics[width=\linewidth]{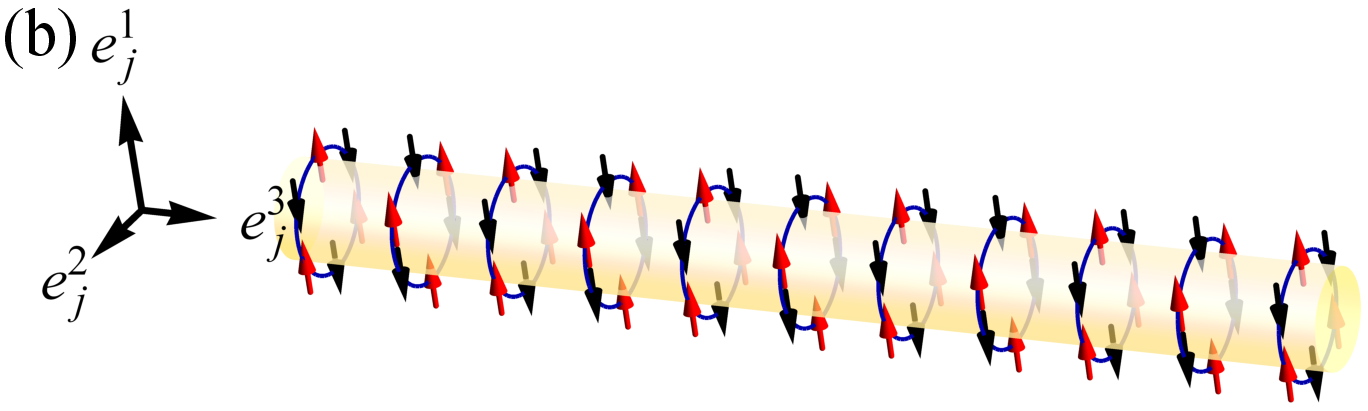}
\caption{A sketch of the antiferromagnetic nuclear spin helix in the (a) original $(\hat{x},\hat{y},\hat{z})$ and (b) rotated $(\hat{e}_{j}^{1},\hat{e}_{j}^{2},\hat{e}_{j}^{3})$ coordinates. The black and red arrows indicate the spins on the sublattice sites A and B, respectively. For simplicity, we do not plot the actual honeycomb lattice here.}
\label{Fig:order}
\end{figure}

We adopt the standard helical ansatz, generalized for the antiferromagnetic correlation between the two sublattices,~\cite{Kornich:2015} 
\begin{equation}
\tilde{{\bf I}}_{\alpha}(r_{j})= \alpha N_{\perp}I \left[
\cos(2k_{F}r_{j})\hat{x} + \sin(2k_{F}r_{j})\hat{y} \right].
\label{Eq:AFHorder}
\end{equation} 
As demonstrated below, this order forms the Neel order in a rotated basis. Although in a conventional antiferromagnet (i.e. Heisenberg antiferromagnet) the Neel order is not the true ground state, it provides a consistent basis for the spin-wave analysis.~\cite{VanKranendonk:1958} We will make sure this is a legitimate choice here by checking the stability of the magnon Hamiltonian. With the order in Eq.~(\ref{Eq:AFHorder}), the nuclear spins are antiferromagnetically aligned on the atomic length scale, whereas they slowly rotate around the helical axis ($z$ direction) on the length scale of $\pi/k_{F}$, as sketched in Fig.~\ref{Fig:order}. 

To derive the magnon spectrum of the antiferromagnetic helix, we rotate the spin axes~\cite{Simon:2008,Braunecker:2009a,Braunecker:2009b} such that in the local basis $(\hat{e}_{j}^{1},\hat{e}_{j}^{2},\hat{e}_{j}^{3})$ the nuclear spin model is mapped onto the Heisenberg antiferromagnet. We write
\begin{eqnarray}
\tilde{{\bf I}}_{\alpha}(r_{j}) &=& \tilde{I}_{\alpha}^{x}(r_{j}) \hat{x} + \tilde{I}_{\alpha}^{y}(r_{j}) \hat{y} + \tilde{I}_{\alpha}^{z}(r_{j}) \hat{z} \nonumber\\
&=& \tilde{I}_{\alpha}^{1}(r_{j}) \hat{e}_{j}^{1} + \tilde{I}_{\alpha}^{2}(r_{j}) \hat{e}_{j}^{2} + \tilde{I}_{\alpha}^{3}(r_{j}) \hat{e}_{j}^{3},
\end{eqnarray}
where the unit vectors in the original and new basis are related by a rotation around the helical axis
\begin{eqnarray}
\left(
\begin{array}{c}
\hat{e}_{j}^{1} \\
\hat{e}_{j}^{2} \\
\hat{e}_{j}^{3}
\end{array}
\right) &=& \left(
\begin{array}{ccc}
\cos(2k_{F} r_{j}) & -\sin(2k_{F} r_{j}) & 0\\
\sin(2k_{F} r_{j}) & \cos(2k_{F} r_{j}) & 0\\
0 & 0 & 1
\end{array}
\right)
\left(
\begin{array}{c}
\hat{x} \\
\hat{y}\\
\hat{z}
\end{array}
\right).\hspace{0.25in} 
\end{eqnarray}
In the rotated basis, the Neel order forms the staggered spin orientation along the $\hat{e}_{j}^{1}$ direction, i.e. $\tilde{{\bf I}}_{\alpha}(r_{j}) = \alpha N_{\perp} I \hat{e}_{j}^{1}$, as sketched in Fig.~\ref{Fig:order}.

The RKKY Hamiltonian, Eq.~(\ref{Eq:RKKY_real}), becomes
\begin{eqnarray}
\mathcal{H}_{\textrm{RKKY}}
&=& \frac{1}{N_{\perp}^2} \sum_{i,j,\alpha,\beta} \sum_{\tilde{\mu}\tilde{\nu}} \tilde{I}_{\alpha}^{\tilde{\mu}} (r_{i}) \tilde{J}_{\alpha\beta}^{\tilde{\mu}\tilde{\nu}}(r_i-r_j)  \tilde{I}_{\beta}^{\tilde{\nu}} (r_{j}),
\label{Eq:RKKY_rot}
\end{eqnarray}
where $\tilde{\mu},\tilde{\nu}=1,2,3$ in spin space and the nonzero components of $\tilde{J}_{\alpha\beta}^{\tilde{\mu}\tilde{\nu}}(r_i-r_j)$ are
\begin{subequations}
\label{Eq:comp_rot}
\begin{eqnarray}
\tilde{J}_{\alpha\beta}^{11}(r_i-r_j) &=& \tilde{J}_{\alpha\beta}^{22}(r_i-r_j) \nonumber\\ 
 &=& J_{\alpha\beta}^{x}(r_i-r_j)  \cos\left[2k_{F} (r_{i}-r_{j})\right], \\
\tilde{J}_{\alpha\beta}^{12}(r_i-r_j) &=& -\tilde{J}_{\alpha\beta}^{21}(r_i-r_j) \nonumber\\
 &=& J_{\alpha\beta}^{x}(r_i-r_j)  \sin\left[2k_{F} (r_{i}-r_{j})\right], \\
\tilde{J}_{\alpha\beta}^{33}(r_i-r_j) &=& J_{\alpha\beta}^{z}(r_i-r_j). 
\end{eqnarray}
\end{subequations}

We now introduce the Holstein-Primakoff transformation for the antiferromagnet,~\cite{VanKranendonk:1958}
\begin{subequations}
\label{Eq:HP_xfmn}
\begin{eqnarray}
\tilde{I}_{A}^{1}(r_{j}) &=& N_{\perp}I - a_{j}^{\dagger} a_{j}, \\
\tilde{I}_{A}^{2}(r_{j}) &=& \sqrt{\frac{N_{\perp}I}{2}} \left( a_{j}^{\dagger} + a_{j} \right), \\
\tilde{I}_{A}^{3}(r_{j}) &=& \sqrt{\frac{N_{\perp}I}{2}} \frac{1}{i} \left( -a_{j}^{\dagger} + a_{j} \right), \\
\tilde{I}_{B}^{1}(r_{j}) &=& -N_{\perp}I + b_{j}^{\dagger} b_{j}, \\
\tilde{I}_{B}^{2}(r_{j}) &=& \sqrt{\frac{N_{\perp}I}{2}} \left( b_{j}^{\dagger} + b_{j} \right), \\
\tilde{I}_{B}^{3}(r_{j}) &=& \sqrt{\frac{N_{\perp}I}{2}} \frac{1}{i} \left( b_{j}^{\dagger} - b_{j} \right),
\end{eqnarray}
\end{subequations}
where the higher order terms in $O(\frac{1}{N_{\perp}I})$ have been neglected.  

Using Eqs.~(\ref{Eq:comp_rot}) and (\ref{Eq:HP_xfmn}) in Eq.~(\ref{Eq:RKKY_rot}) and performing the Fourier transform, 
\begin{subequations}
\begin{eqnarray}
a_{q} &=& \frac{1}{\sqrt{N}} \sum_{j} e^{-iqr_{j}} a_{j},\\
b_{q} &=& \frac{1}{\sqrt{N}} \sum_{j} e^{iqr_{j}} b_{j},
\end{eqnarray}
\end{subequations}
we obtain the magnon Hamiltonian in momentum space,
\begin{eqnarray}
\mathcal{H}_{\textrm{magnon}} &=& 
\frac{I}{2N_{\perp}} \sum_{q} 
 \Psi_{\textrm{magnon}}^{\dagger}(q)
\mathcal{D}(q)
\Psi_{\textrm{magnon}}(q),
\label{Eq:magnon}
\end{eqnarray}
where $\Psi_{\textrm{magnon}}^{\dagger}(q) = \left( a_{q}^{\dagger}, a_{-q}^{\dagger},  b_{q}^{\dagger}, b_{-q}^{\dagger}, a_{q}, a_{-q}, b_{q}, b_{-q} \right)$. The 8-by-8 symmetric matrix $\mathcal{D}(q)$ is
\begin{eqnarray}
\mathcal{D}(q) &\equiv& \left(
\begin{array}{cc}
\mathcal{A}(q) & \mathcal{B}(q) \\
\mathcal{B}(q) & \mathcal{A}(q) \\
\end{array}
\right), 
\end{eqnarray} 
where the 4-by-4 block matrices $\mathcal{A}(q)$ and $\mathcal{B}(q)$ are defined as
\begin{eqnarray}
\mathcal{A}(q) &\equiv& 
\left( 
\begin{array}{cccc}
h_{3}(q) &0 & 0 & h_{2}(q) \\
0 & h_{3}(q) & h_{2}(q) & 0 \\
0 &  h_{2}(q) & h_{3}(q) & 0 \\
h_{2}(q) & 0 & 0 & h_{3}(q) 
\end{array}
\right), \\
\mathcal{B}(q) &\equiv& 
\left( 
\begin{array}{cccc}
0 & -h_{2}(q) & h_{1}(q) & 0 \\
-h_{2}(q) & 0 & 0 &  h_{1}(q) \\
h_{1}(q) & 0 & 0 &  -h_{2}(q) \\
0 & h_{1}(q) & -h_{2}(q) & 0  
\end{array}
\right),
\end{eqnarray}
with
\begin{subequations} 
\begin{eqnarray}
h_{1}(q) &\equiv& \frac{1}{4} \left[ J_{AB}^{x}(q-2k_{F}) +  J_{AB}^{x}(q+2k_{F}) + 2J_{AB}^{z}(q) \right], \nonumber\\
~\\
h_{2}(q) &\equiv& \frac{1}{4} \left[ J_{AB}^{x}(q-2k_{F}) +  J_{AB}^{x}(q+2k_{F}) - 2J_{AB}^{z}(q) \right], \nonumber\\
~\\
h_{3}(q) &\equiv& 2J_{AB}^{x}(2k_{F})-h_{1}(q).
\end{eqnarray}
\end{subequations}
One can check that $\mathcal{D}(q)$ is positive definite, which ensures the stability of the nuclear spin order and justifies our ansatz for the antiferromagnetic helix.~\cite{vanHemmen:1980} 

\begin{figure}[htb]
\centering
\includegraphics[width=\linewidth]{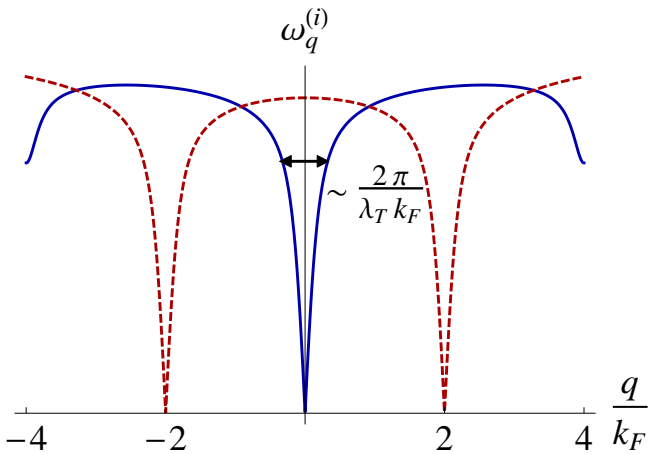}
\caption{Magnon spectrum of the antiferromagnetic helix. The parameters used here are the same as in Fig.~\ref{Fig:RKKY}. The blue solid and red dashed lines correspond to the $\omega_{q}^{(1)}$, and $\omega_{q}^{(2)}$ energy bands, respectively. As in Fig.~\ref{Fig:RKKY}, here we use an unrealistically high temperature to illustrate the dips in the spectrum. The region of the linear dispersion is given by the width of the RKKY peaks $\sim\frac{2\pi}{\lambda_{T}k_{F}}$, which is much narrower for realistic temperatures.}
\label{Fig:magnon}
\end{figure}

The excitation spectrum of Eq.~(\ref{Eq:magnon}) is given by twice the positive eigenvalues of the matrix
\begin{eqnarray}
\left(
\begin{array}{cc}
\mathcal{A}(q) & \mathcal{B}(q) \\
-\mathcal{B}(q) & -\mathcal{A}(q)
\end{array}
\right). 
\end{eqnarray}
Diagonalization gives two magnon bands,~\cite{vanHemmen:1980}
\begin{eqnarray}
\hbar \omega_{q}^{(1)} 
&=& \frac{I}{N_{\perp}} \sqrt{2J_{AB}^{x}(2k_{F})} \nonumber \\
&\times&  \sqrt{ 2 J_{AB}^{x}(2k_{F}) - J_{AB}^{x}(q-2k_{F}) - J_{AB}^{x}(q+2k_{F}) } , \nonumber\\
~\\
\hbar \omega_{q}^{(2)} &=& 
\frac{2I}{N_{\perp}} \sqrt{J_{AB}^{x}(2k_{F}) \left[ J_{AB}^{x}(2k_{F}) - J_{AB}^{z}(q) \right]},
\end{eqnarray}
which are shown in Fig.~\ref{Fig:magnon}. 

In Fig.~\ref{Fig:magnon}, one can see that there are zero-energy excitations at $q=0,\pm 2k_{F}$. These Goldstone modes are protected by the symmetries in the system; namely, the rotational symmetry of the nuclear spins around the helical axis and the rotation of the helical axis itself.~\cite{Klinovaja:2013b,Meng:2014} Around these nodes, the low-energy magnon spectrum exhibits linear dispersions. In the (nonhelical) Heisenberg model, the low-energy dispersion of the magnons is quadratic in the ferromagnetic case, whereas it becomes linear in the antiferromagnetic case.~\cite{VanKranendonk:1958,Ashcroft:1976} Interestingly we find that for one-dimensional helical systems, both locally ferromagnetic~\cite{Braunecker:2009a,Braunecker:2009b} and antiferromagnetic orders contain low-energy magnons with linear dispersions, 
but differences in scaling coefficients, as discussed in Sec.~\ref{SubSec:Tc}.

\subsection{Transition temperature of the nuclear spin order without the feedback \label{SubSec:Tc}}

We now estimate the transition temperature by considering the fluctuations due to the magnons, which reduce the sublattice magnetization. From the magnon spectrum one can see that the long-wavelength magnons ($q\approx 0$) and their two replicas at $q\approx \pm 2 k_F$ have the smallest energy, with linear dispersion. In an infinite system, such excitations destroy order at any finite temperature. Namely, whereas the original Mermin-Wagner theorem~\cite{Mermin:1966} and its extension~\cite{Bruno:2001} for oscillatory exchange interaction in free electron gas do not apply to this system, an extension of the theorem for a more generic Hamiltonian, including the electron-electron interaction, rules out any spontaneous orders at finite temperatures in the thermodynamic limit.~\cite{Loss:2011} 
However, in a finite system with length $L$, the lowest allowed momentum is given by $q_{1}=\frac{\pi}{L}$ and the values of the excitation momenta are discrete (not continuous), so that an order may be established in principle.
This finite-size-induced finite energy of long-wavelength excitations corresponds to a gap of the zero-energy Goldstone modes.

In addition, here the peak in the susceptibility is so sharp that the region where magnons can be considered long-wavelength (that is, having linear dispersion) is extremely narrow. 
In fact, from the analysis in Ref.~\citenum{Meng:2014} it follows that for sample sizes realistic for nanowires and nanotubes, the long-wavelength magnons are completely negligible and the transition temperature can be obtained by considering only the contribution from the short-wavelength magnons. If $q_{1}>\frac{\pi}{\lambda_{T}}$ (equivalently $k_{B}T<\frac{\hbar v_{F}}{L}$, which in our case is satisfied for any realistic length), then these magnons have approximately a momentum-independent energy, $\hbar \omega_{m}$, of order $|J_{\alpha\beta}^{x}(q=2k_{F})|$, a property which makes the transition temperature calculation analytically tractable. 
Namely, the temperature dependence of the magnon occupation can be computed by
\begin{equation}
 N_{\perp} \times \sum_{i=1,2} \sum_{q}{}^{\prime} \frac{1}{e^{\frac{\hbar \omega_{m}}{k_{B}T}}-1},
\label{Eq:mag_occupation}
\end{equation}
where the summation over $i=1,2$ includes both magnon bands, and the prime on the summation denotes that the Goldstone modes are excluded. 
Finally, the prefactor $N_{\perp}$ is required to reflect the $N_{\perp}$ possibilities to flip a spin within a cross section for a short-wavelength magnon.~\cite{Meng:2014}

The order parameter, defined as the $q=2k_{F}$ component of the normalized staggered magnetization, i.e. the normalized sublattice magnetization, can then be expressed as
\begin{eqnarray}
m_{2k_{F}}(T) &=& 1 - \frac{1}{NI} \sum_{i=1,2} \sum_{q}{}^{\prime} \frac{1}{e^{\frac{\hbar \omega_{m}}{k_{B}T}}-1},
\label{Eq:m2kF}
\end{eqnarray}
which equals unity for the fully ordered nuclear spin state, and vanishes for completely disordered phase.

The constant magnon energies, $\hbar \omega_{m} = 2I J_{AB}^{x}(2k_{F},T) /N_{\perp}$, lead to a generalized Bloch law,~\cite{Braunecker:2009b}
\begin{eqnarray}
m_{2k_{F}}(T)
&=& 1 - \left( \frac{T}{T_{0}} \right)^{3-2g_{x}}
\label{Eq:BlochLaw}
\end{eqnarray}
with a non-universal exponent $(3-2g_{x})$ modified by the electron-electron interaction. We also define
\begin{eqnarray}
k_{B}T_{0}
&\approx& \left[ \frac{I^2  A_{0}^2}{N_{\perp}}
\left( \Delta_{a} \right)^{1-2g_{x}} C(g_{x}) \right]^{\frac{1}{3-2g_{x}}},
\label{Eq:T0}
\end{eqnarray} 
where $\Delta_{a}\equiv \hbar v_{F}/a$ is the bandwidth, and
\begin{eqnarray}
C(g)&\equiv& \frac{1}{8} \sin (\pi g) (2\pi)^{2g-4}
\left| 
\frac{ \Gamma\left( 1-g\right) \Gamma\left( \frac{g}{2} \right)}
{\Gamma\left( \frac{2-g}{2} \right)}
\right|^{2}.
\end{eqnarray}
We note that even for the noninteracting limit $g_{x}\rightarrow 1$, the exponent in Eq.~(\ref{Eq:BlochLaw}) is still different from the $T^{\frac{3}{2}}$ law for Heisenberg ferromagnets~\cite{Bloch:1930,VanKranendonk:1958,Ashcroft:1976} or the $T^{2}$ law for Heisenberg antiferromagnets.~\cite{Kubo:1952,Oguchi:1960}
The nanotube radius $R$ has two effects on $T_{0}$: first, as mentioned in Sec.~\ref{SubSec:Hel}, larger $R$ results in less prominent electron-electron interaction, and thus a larger exponent $g_{x}$. Second, larger $R$, which is proportional to $N_{\perp}$ that enters Eq.~(\ref{Eq:T0}), weakens the finite-size effect, so the magnon occupation increases, as indicated in Eq.~(\ref{Eq:mag_occupation}). 
As a result of both of these effects, CNTs with larger $R$ are expected to have a lower transition temperature. In addition, we also note that for CNTs not purely made of $^{13}$C, say with the $^{13}$C concentration of $p$, Eq.~\eqref{Eq:BlochLaw} is valid upon replacement  $A_{0}\rightarrow pA_{0}$ in Eq.~(\ref{Eq:T0}), so that the transition temperature will be reduced as $T_{0} \propto p^{2/(3-2g_{x})}$. 

Using Eq.~(\ref{Eq:T0}) we evaluate $T_{0}\approx 1.9~\mu$K, which is too small for dilution fridge experiments. However, so far we have not included the Overhauser field due to the nuclear spins, which acts back on the conduction electrons and further stabilizes the order. In the next section, we take into account this feedback and estimate how it modifies the transition temperature.

As a self-consistency check, we examine that the energy scale of the RKKY interaction, Eq.~\eqref{Eq:RKKY_real}, does not excess the one set by the original hyperfine Hamiltonian, Eq.~\eqref{Eq:H_hf}. The former energy scale is dominated by the $q=2k_{F}$ peak, Eq.~\eqref{Eq:peak}. Thus, after Fourier transforming Eq.~\eqref{Eq:RKKY_real} into momentum space through $\tilde{I}_{\alpha}^{\mu}(q)=\sum_{j} e^{-iqr_{j}}\tilde{I}_{\alpha}^{\mu}(r_{j})$, we obtain
\begin{eqnarray}
E_{\textrm{RKKY}} &\approx& \frac{1}{NN_{\perp}^2} \left| 
J_{AB}^{x}(q=2k_{F}) \tilde{I}_{A}^{x}(2k_{F})\tilde{I}_{B}^{x}(-2k_{F})
\right| \nonumber\\
& \approx & N J_{AB}^{\mu}(q=2k_{F}) I^2,
\label{Eq:check1}
\end{eqnarray} 
where we keep only the dominant $q=2k_{F}$ component, and $\tilde{I}_{\alpha}^{x}(q=\pm 2k_{F})$ is replaced by its maximal value, $NN_{\perp} I$.
The energy scale of the hyperfine Hamiltonian, on the other hand, can be obtained by considering all electrons are polarized such that their spins locally align with the nuclear spins.~\cite{Braunecker:2009b} This gives 
\begin{eqnarray}
E_{\textrm{hf}} &\approx& \sum_{j,\alpha} \frac{A_{0}}{2} \frac{n_{el}}{n_{I}} \left|\tilde{I}_{\alpha}(r_{j})\right| \nonumber\\
& \approx & N A_{0} I \frac{k_{F} a}{\pi},
\label{Eq:check2}
\end{eqnarray} 
where $n_{el}=(2k_{F}/\pi)|\psi_{\perp,\alpha}|^2 =k_{F}/(\pi^2 R)$ is the area electron density, and the electron spin is included through the factor of $\frac{1}{2}$. Here $n_{I}=N_{\perp}/(\pi Ra)$ is the area nuclei density, introduced in Sec.~\ref{SubSec:Hamiltonian}. In the second line we replace $\tilde{I}_{\alpha}^{\mu}(r_{j})$ by its length, $N_{\perp} I$, and the summation gives a factor of $2N$.
Combining Eqs.~\eqref{Eq:check1} and \eqref{Eq:check2}, we obtain the self-consistency condition,
\begin{eqnarray}
J_{AB}^{\mu}(q=2k_{F}) \le A_{0} \frac{2 k_{F} a}{\pi},
\end{eqnarray} 
where $J_{AB}^{\mu}(q=2k_{F})$ is temperature dependent. The above condition is fulfilled for $T=T_{0}$.

\section{Feedback effects \label{Sec:feedback}}

\subsection{Overhauser field from the nuclear spin order\label{SubSec:Overhauser}}

Since $A_{0}\ll \epsilon_{F}$, the characteristic time scales of the slow nuclear and fast electron dynamics can be considered to be decoupled. Therefore, we can treat the nuclear spin order as a static order, which induces a static spatially oscillating Overhauser field that acts back on the electrons. 
Including the antiferromagnetic helix in the hyperfine coupling terms, we obtain 
\begin{eqnarray}
\mathcal{H}_{\textrm{fb}} &=& \frac{A_{0}}{N_{\perp}} \sum_{\alpha,j} {\bf S}_{\alpha}(r_{j}) \cdot \left\langle {\bf \tilde{I}}_{\alpha}(r_{j}) \right\rangle.
\end{eqnarray}
The antiferromagnetic helix with $q=2k_{F}$ gives
\begin{eqnarray}
\left\langle {\bf \tilde{I}}_{\alpha}(r_{j}) \right\rangle &=& \alpha N_{\perp} I m_{2k_{F}} \left[ \cos \left( 2k_{F}r_{j} \right) \hat{x} + \sin \left( 2k_{F}r_{j} \right) \hat{y}
\right]. \nonumber\\
\end{eqnarray}
In the continuum limit we then have
\begin{eqnarray}
\mathcal{H}_{\textrm{fb}} &=& \sum_{\alpha} \int dr \; 
{\bf B}_{Ov,\alpha}(r) \cdot {\bf S}_{\alpha}(r),
\end{eqnarray}
where the nuclear Overhauser field is defined as
\begin{eqnarray}
{\bf B}_{Ov,\alpha}(r) = \alpha B_{Ov} \left[ \cos \left( 2k_{F}r \right) \hat{x} + \sin \left( 2k_{F}r \right) \hat{y}
\right] 
\end{eqnarray}
with $B_{Ov} \equiv A_{0} I m_{2k_{F}}$. 
The summation over $\alpha$ eliminates the intervalley back scattering terms, Eq.~(\ref{Eq:spin_operator_inter}), so only the intravalley back scattering terms, Eq.~(\ref{Eq:spin_operator_intra}), enter $\mathcal{H}_{\textrm{fb}}$. 

With Eqs.~(\ref{Eq:SDW_operator})--(\ref{Eq:SDW_z}), the feedback Hamiltonian can be written as
\begin{eqnarray}
\mathcal{H}_{\textrm{fb}}
 &=& \frac{ B_{Ov} }{2\pi a} \sum_{\gamma} \int dr \left[
\cos \left( \phi_{cS} + \gamma \phi_{cA} + \theta_{sS} + \gamma \theta_{sA} 
\right) \right. \nonumber \\
&& \left. \hspace{0.3in} + \cos \left( \phi_{cS} + \gamma \phi_{cA} - \theta_{sS} - \gamma \theta_{sA} -4k_{F} r \right) \right], \nonumber\\
\end{eqnarray}
where we neglected the forward scattering part because it has no influence.~\cite{Braunecker:2009a,Braunecker:2009b}
The cosine in the second term oscillates except for the commensurate case, $2k_{F}a=2\pi \times \textrm{integer}$. The commensurate case corresponds to an unrealistic gate-tuning, so we assume the system is incommensurate and drop the second cosine term.~\cite{Braunecker:2009a,Braunecker:2009b} Consequently we have the sine-Gordon term
\begin{eqnarray}
\mathcal{H}_{\textrm{fb}} 
 &\approx& \frac{ B_{Ov} }{2\pi a} \sum_{\gamma} \int dr \left[
\cos \left( \phi_{cS} + \gamma \phi_{cA} + \theta_{sS} + \gamma \theta_{sA} 
\right) \right], \nonumber\\
\label{Eq:H_s-G}
\end{eqnarray}
which is renormalization-group (RG) relevant in the interacting system as discussed in Sec.~\ref{SubSec:RG}. Therefore, it will gap out the $(\phi_{cS} + \gamma \phi_{cA} + \theta_{sS} + \gamma \theta_{sA})$ modes, but leave $(\phi_{cS} + \gamma \phi_{cA} - \theta_{sS} - \gamma \theta_{sA})$ modes gapless, which can still effectively mediate the RKKY interaction. 

Before analyzing the Overhauser field due to the antiferromagnetic nuclear spin helix, let us come back to the intervalley back scattering terms of the spin operator, Eq.~(\ref{Eq:spin_operator_inter}), which would have led to a sublattice-independent ferromagnetic coupling $J_{AA}^{\mu}(q)=J_{AB}^{\mu}(q)<0$, and hence a locally ferromagnetic helical order with $q=2(k_{F}-\gamma k_{v})$,
\begin{eqnarray}
\left\langle {\bf \tilde{I}}_{\textrm{fm},\alpha}(r_{j}) \right\rangle &=& \alpha N_{\perp} I m_{2(k_{F}-\gamma k_{v})} 
\left\{ \cos \left[ 2(k_{F}-\gamma k_{v})r_{j} \right] \hat{x} \right.
\nonumber\\
&& \hspace{0.3in}
\left. + \sin \left[ 2(k_{F}-\gamma k_{v})r_{j} \right] \hat{y}
\right\}.
\end{eqnarray}
The corresponding feedback Hamiltonian is 
\begin{eqnarray}
\mathcal{H}_{\textrm{fb,inter}} 
 &=& 
\frac{A_{0}}{N_{\perp}} \sum_{\alpha,j} {\bf S}_{b,\textrm{inter},\alpha}(r_{j}) \cdot \left\langle {\bf \tilde{I}}_{\textrm{fm},\alpha}(r_{j}) \right\rangle \nonumber \\
&\approx& 
\frac{ B'_{Ov}  }{2\pi a} \sum_{\gamma} \int dr \left[
\cos \left( \phi_{cS} - \gamma \phi_{sA} - \theta_{sS} + \gamma \theta_{cA} 
\right) \right], \nonumber\\
\end{eqnarray}
where $B'_{Ov}\equiv A_{0} I m_{2(k_{F}-\gamma k_{v})}$, and the oscillating terms are omitted again. Here the cosine terms will gap out the $(\phi_{cS} - \gamma \phi_{sA} - \theta_{sS} + \gamma \theta_{cA})$ modes, but leave the $(\phi_{cS} + \gamma \phi_{sA} + \theta_{sS} + \gamma \theta_{cA})$ modes gapless. However, while both of the $(\phi_{cS} + \gamma \phi_{sA} + \theta_{sS} + \gamma \theta_{cA})$ modes for $\gamma=\pm$ may mediate the ferromagnetic RKKY interaction, they generate different extrema at $q=2(k_{F} \mp k_{v})$, respectively. On the other hand, for the intravalley back scattering, both $\gamma=\pm$ valleys produce the same extremum at $q=2k_{F}$, so the absolute value of the magnitude of the RKKY interaction at $q=2k_{F}$ will be twice larger than the ones at $q=2(k_{F} -\gamma k_{v})$. 

Comparing the two scenarios, the energy gains by forming these two nuclear spin orders are different because of the different peak heights, even though both the intervalley and intravalley back scattering terms lead to peaks in the RKKY interaction. Consequently, the ground state favors the antiferromagnetic helix with $q=2k_{F}$ to minimize the energy.
In addition, the gapping of half of the conduction electron modes reduces the conductance by a factor of 2, as predicted in Refs.~\citenum{Braunecker:2009a,Braunecker:2009b} for materials with no valley degrees of freedom, which may have been observed in GaAs nanowires.~\cite{Scheller:2014}

\subsection{Renormalized Overhauser field \label{SubSec:RG}}

\begin{figure}[htb]
\centering
\includegraphics[width=\linewidth]{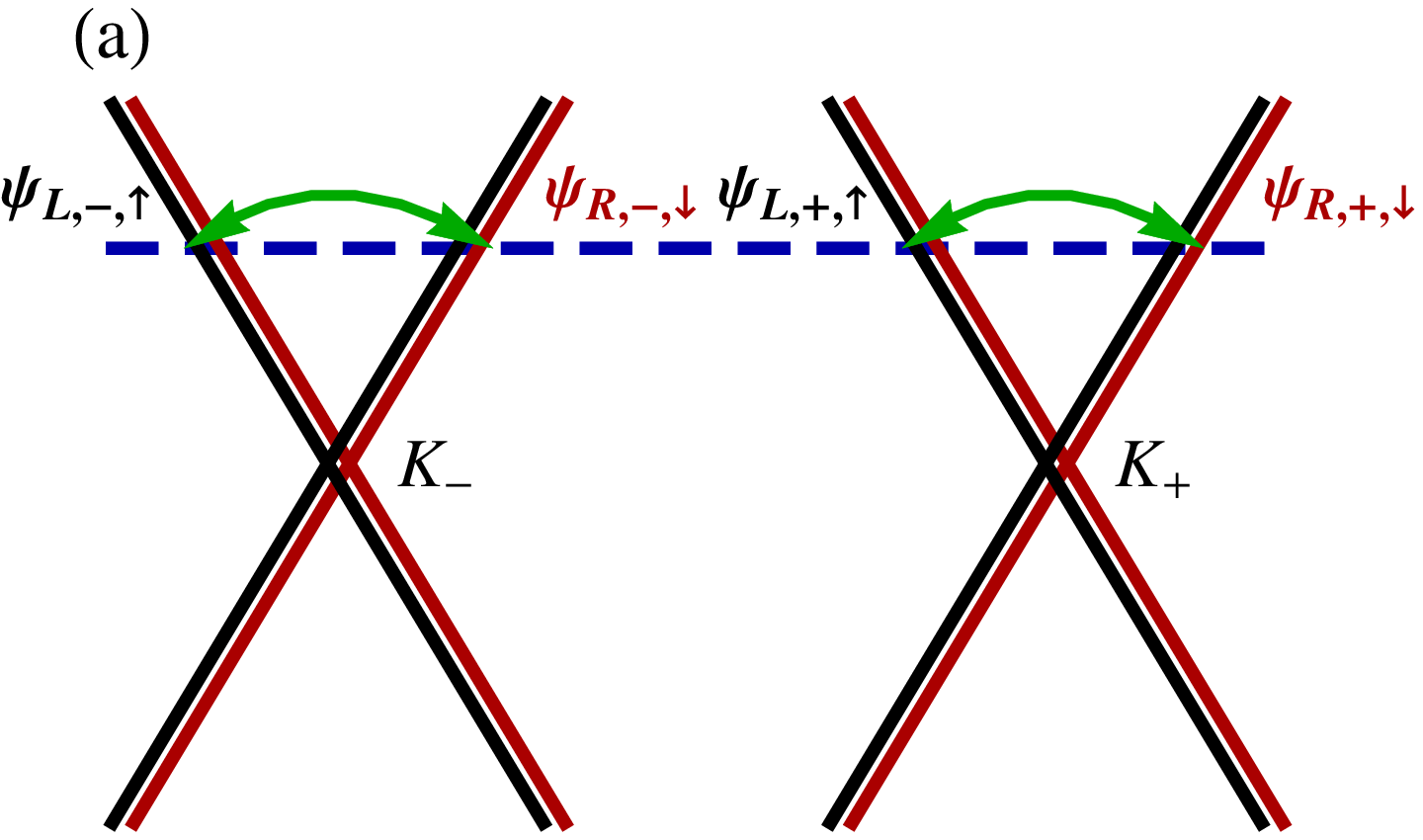}
\includegraphics[width=\linewidth]{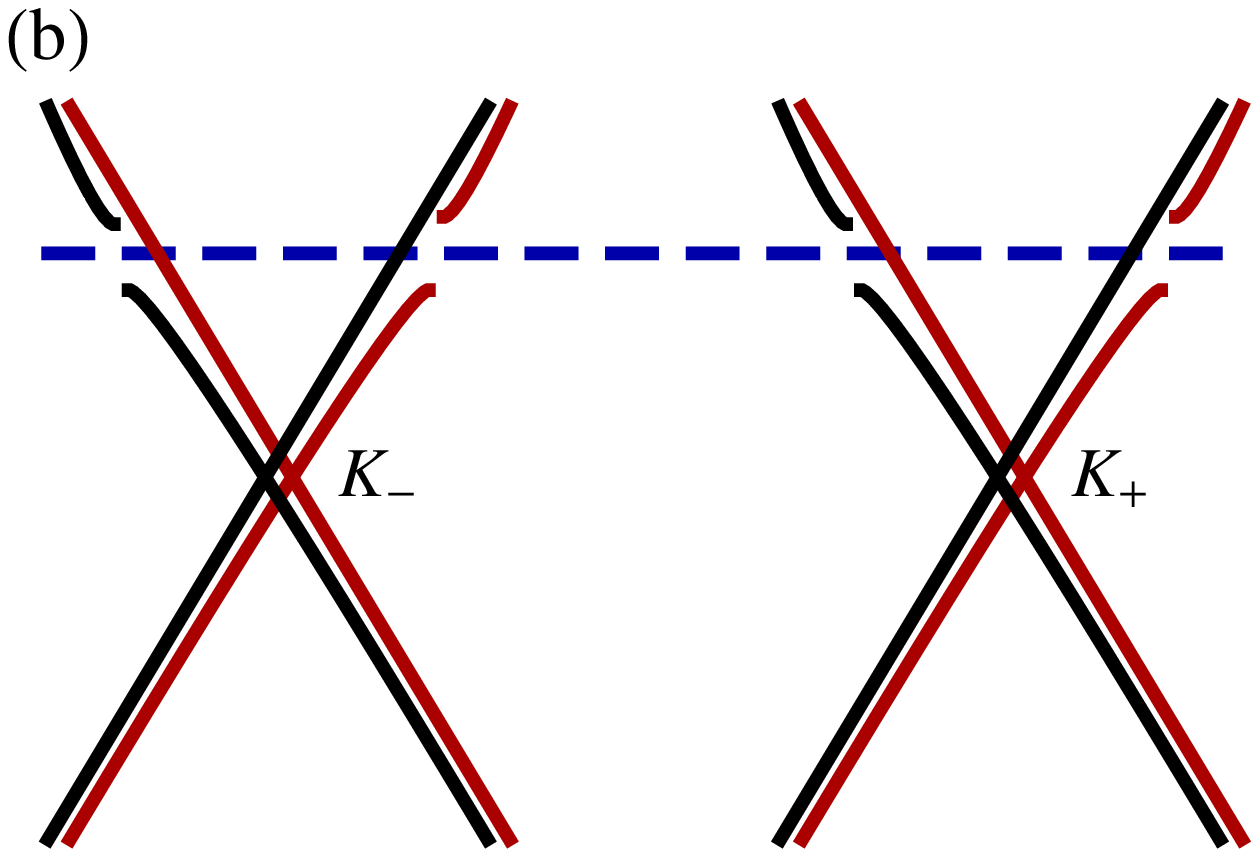}
\caption{The antiferromagnetic helix corresponds to the intravalley back scattering processes (a), gapping out half of conduction electrons (b). The up and down spins are marked in black and red colors, respectively. The chemical potential is plotted with the blue dashed lines. The green arrows describe the scattering processes. The dispersions for up and down spins are slightly shifted for clarity.}
\label{Fig:helix_scattering}
\end{figure}

Based on the analysis in Sec.~\ref{SubSec:Overhauser}, the system will organize the nuclear spins to maximize the $m_{2k_{F}}$ component with antiferromagnetic helix to lower the energy. Therefore, from now on we drop the intervalley back scattering contribution to the feedback effects, and consider only the antiferromagnetic helix due to the intravalley back scattering terms in Eq.~(\ref{Eq:H_s-G}). 
In terms of the right and left moving particles, the feedback term describes the $(L,\uparrow)\leftrightarrow(R,\downarrow)$ scattering within each valley,
\begin{equation}
\psi_{L,\gamma,\uparrow}^{\dagger} \psi_{R,\gamma,\downarrow} + \psi_{R,\gamma,\downarrow}^{\dagger} \psi_{L,\gamma,\uparrow},
\end{equation}
as illustrated in Fig.~\ref{Fig:helix_scattering}.
Notice that this is a consequence of the choice of helicity in Eq.~(\ref{Eq:AFHorder}); if the other helicity is chosen, namely, $\left[
\cos(2k_{F}r_{j})\hat{x} - \sin(2k_{F}r_{j})\hat{y} \right]$, then the antiferromagnetic helix will correspond to the $(L,\downarrow)\leftrightarrow(R,\uparrow)$ scattering, which gaps out different spin subbands of conduction electrons. Even though half of the conduction electrons are gapped by the nuclear spin order, the feedback strongly renormalizes the other half of the electrons, leading to stronger effective electron-electron interaction, as can be seen below. 

To proceed we define a new set of bosonic fields,
\begin{eqnarray}
\Phi_{\gamma}^{\pm} &\equiv& \frac{1}{2} \left[ 
\pm \left(\phi_{cS} + \gamma \phi_{cA} \right) 
+ \left(\theta_{sS} + \gamma \theta_{sA} \right) \right], 
\label{Eq:newfield1}\\
\Theta_{\gamma}^{\pm} &\equiv& \frac{1}{2} \left[ 
\left(\phi_{sS} + \gamma \phi_{sA} \right) 
\pm  \left(\theta_{cS} + \gamma \theta_{cA} \right) \right],
\label{Eq:newfield2}
\end{eqnarray}
which satisfy the commutation relations~\cite{Giamarchi:2003,Bruus:2004}
\begin{eqnarray}
\left[ \Phi_{\gamma}^{\pm}(r_{1}), \Theta_{\gamma}^{\pm}(r_{2}) \right] &=& i \frac{\pi}{2} \textrm{sign}(r_{2}-r_{1}).
\end{eqnarray}
In terms of the new fields, the feedback Hamiltonian, 
Eq.~\eqref{Eq:H_s-G}, is
\begin{eqnarray}
\mathcal{H}_{\textrm{fb}} &=& 
\frac{ B_{Ov} }{2\pi a} \sum_{\gamma} \int dr 
\cos \left( 2 \Phi_{\gamma}^{+} \right),
\label{Eq:H_fb}
\end{eqnarray}
and the electronic Hamiltonian, Eq.~\eqref{Eq:LL}, becomes
\begin{eqnarray}
\mathcal{H}_{\textrm{el}} &\approx& 
\sum_{\gamma} \int \frac{\hbar dr}{2\pi} 
\left\{
\tilde{u} \tilde{K}
\left[ \left(\triangledown \Theta_{\gamma}^{+}\right)^2 + \left(\triangledown \Theta_{\gamma}^{-}\right)^2 \right] \right. \nonumber\\ 
&& \hspace{0.65in} \left.
+ \frac{\tilde{u}}{\tilde{K}}
\left[ \left(\triangledown \Phi_{\gamma}^{+}\right)^2 + \left(\triangledown \Phi_{\gamma}^{-}\right)^2 \right] \right\}, 
\label{Eq:LL_modified}
\end{eqnarray}
where the cross terms, such as $\left(\triangledown \Phi_{\gamma}^{\pm} \right) \left(\triangledown \Phi_{\bar{\gamma}}^{\pm} \right)$ and $\left(\triangledown \Theta_{\gamma}^{\pm} \right) \left(\triangledown \Theta_{\bar{\gamma}}^{\pm} \right)$, have been neglected because they are marginal and less important than the relevant cosine terms.~\cite{Braunecker:2009b,Meng:2013} Here the modified velocity and Luttinger liquid parameter in the presence of the feedback are given by
\begin{eqnarray}
\tilde{u} &\equiv& \frac{1}{4}\left[ \sum_{P,P'} \left( u_{c P} K_{c P} + \frac{u_{s P}}{K_{s P}} \right) 
\left( u_{s P'} K_{s P'} + \frac{u_{c P'}}{K_{c P'}} \right) \right]^{\frac{1}{2}}, \nonumber\\
\tilde{K} &\equiv& \left[ \frac{\sum_{P} \left( u_{c P} K_{c P} + \frac{u_{s P}}{K_{s P}} \right)}
{\sum_{P'} \left( u_{s P'} K_{s P'} + \frac{u_{c P'}}{K_{c P'}} \right)}
\right]^{\frac{1}{2}}.
\label{Eq:k_tilde}
\end{eqnarray}
For the spin isotropic systems, $K_{sS}=K_{sA}=1$, we have $\tilde{u}=v_{F}/\tilde{K}$. For noninteracting systems, $\tilde{K}=1$ and $\tilde{u}=v_{F}$ are recovered. For CNTs, we have $\tilde{K} \approx 0.38$ and $\tilde{u}\approx 2.6 v_{F}$.

Eqs.~\eqref{Eq:H_fb} and \eqref{Eq:LL_modified} state that the $(\Phi_{\gamma}^{+},\Theta_{\gamma}^{+})$ and $(\Phi_{\gamma}^{-},\Theta_{\gamma}^{-})$ sectors are decoupled, with the former described by a sine-Gordon Hamiltonian, and the latter by a free bosonic one. To analyze the sine-Gordon Hamiltonian, we define a dimensionless coupling constant, $\tilde{y}(l)\equiv B_{Ov}(l)/\tilde{\Delta}_{a}(l)$ with $\tilde{\Delta}_{a}(l) \equiv \hbar \tilde{u}/\xi(l)$ and correlation length $\xi(l) \equiv ae^{l}$. 
Then, we obtain the RG flow equation for $\tilde{y}(l)$,~\cite{Giamarchi:2003}
\begin{equation}
\frac{d \tilde{y}(l)}{dl} = \left( 2 - \tilde{K} \right) \tilde{y}(l),
\label{Eq:RG}
\end{equation} 
where $l$ is the cutoff length scale. In the systems under consideration, we always have $2 - \tilde{K} > 0$, so $\tilde{y}(l)$ grows under the RG flow as 
\begin{equation}
\tilde{y}(l) = \tilde{y}(0) e^{\left( 2 - \tilde{K} \right)l},
\end{equation}
and the cosine term is relevant. The renormalized Overhauser field is then 
\begin{equation}
B_{Ov}^{*} = B_{Ov} \left( \frac{\xi}{a} \right)^{( 1 - \tilde{K} )}.
\end{equation}

The RG flow will stop when $\xi$ exceeds the system size $L$, the thermal length $\lambda_{T}$, or at a scale $l^{*}$, where the coupling constant becomes of order 1, $y(l^{*})\approx 1$, which gives 
\begin{eqnarray}
\xi(l^{*}) &\equiv& a e^{l^{*}} = a\left( \frac{IA_{0}}{\tilde{\Delta}_{a}} \right)^{-\frac{1}{2-\tilde{K}}},
\end{eqnarray}
with $\tilde{\Delta}_{a} \equiv \tilde{\Delta}_{a}(l=0) = \hbar \tilde{u}/a$.
The correlation length is determined by the smallest scale at which any of the above conditions is reached,
\begin{equation}
\xi=\textrm{min} \left\{ L,\tilde{\lambda}_{T} \equiv \frac{\hbar \tilde{u}}{k_{B}T}, \xi(l^{*}) \right\}.
\end{equation}
In CNTs, a typical system size is of order $L=1~\mu$m. In addition, $\tilde{\lambda}_{T}(T=10~\textrm{mK})\approx 1.6$~mm, and $\xi(l^{*}) \approx 1.8~\mu$m, so $L \apprle \xi(l^{*}) \ll \lambda_{T}$ and we obtain $\xi=L=1~\mu$m. 

The renormalized hyperfine coupling constant, to which the Overhauser field is proportional, is then
\begin{equation}
A^{*} = A_{0} \left( \frac{\xi}{a} \right)^{( 1 - \tilde{K} )}.
\end{equation}
For the noninteracting systems $\tilde{K}=1$, and the coupling is the bare one. For CNTs, we get $|A^{*}|\approx 180 |A_{0}| \approx 1.1~$meV. Importantly, the renormalization is stronger than the one obtained within the one-band description ($A^{*}\approx 22~\mu$eV)~\cite{Braunecker:2009a,Braunecker:2009b} because of the smaller exponent $\tilde{K}$ here. 
With the renormalization we still have $A^{*} \ll \epsilon_{F} \approx 0.1$~eV, so the Schrieffer-Wolff transformation remains well defined.
The gap due to the antiferromagnetic helix can be obtained from the RG analysis,~\cite{Giamarchi:2003}
\begin{eqnarray}
\Delta_{m} = \tilde{\Delta}_{a} \left( \frac{IA_{0}}{\tilde{\Delta}_{a}} \right)^{\frac{1}{2-\tilde{K}}},
\end{eqnarray}
which leads to a gap of $\Delta_{m}\approx 0.77~$meV for our parameters.

\subsection{Transition temperature in the presence of the feedback \label{SubSec:Tc_fb}}

In this section, we include the feedback into the Hamiltonian, and compute the spin susceptibility in the presence of the Overhauser field. The modified RKKY interaction is proportional to the modified static spin susceptibility, which is now evaluated with the modified electronic Hamiltonian, $\mathcal{H}_{\textrm{el}} + \mathcal{H}_{\textrm{fb}}$, where $\mathcal{H}_{\textrm{el}}$ and $\mathcal{H}_{\textrm{fb}}$ are given by Eqs.~(\ref{Eq:LL_modified}) and (\ref{Eq:H_fb}), respectively.

The modified correlation functions $\tilde{\chi}_{AA}^{>,\mu}(r)=-\tilde{\chi}_{AB}^{>,\mu}(r)$ are
\begin{widetext}
\begin{eqnarray}
\tilde{\chi}_{AA}^{>,x}(\tilde{r}) &=& \tilde{\chi}_{AA}^{>,y}(\tilde{r}) \nonumber\\
&=& \frac{-i \cos(2k_{F}r)}{2(4\pi a)^2} 
\sum_{\gamma}
\left\{ \left\langle 
e^{i \sqrt{2} \Phi_{\gamma}^{+}(\tilde{r}) } 
e^{-i \sqrt{2} \Phi_{\gamma}^{+}(0)}
\right\rangle 
+ \left\langle
e^{i \sqrt{2} \Phi_{\gamma}^{-}(\tilde{r}) } 
e^{- i \sqrt{2} \Phi_{\gamma}^{-}(0) }
\right\rangle \right\},
\label{Eq:chi_greater_fb} \\
\tilde{\chi}_{AA}^{>,z}(\tilde{r}) 
&=& \frac{-i \cos(2k_{F}r)}{2(4\pi a)^2} 
\sum_{\gamma}
\left\{ 
\left\langle 
e^{\frac{i}{\sqrt{2}}  
\left[ \Phi_{\gamma}^{+}(\tilde{r}) -\Phi_{\gamma}^{-}(\tilde{r}) 
+ \Theta_{\gamma}^{+} (\tilde{r}) + \Theta_{\gamma}^{-} (\tilde{r}) \right]}
e^{-\frac{i}{\sqrt{2}}  
\left[ \Phi_{\gamma}^{+}(0) - \Phi_{\gamma}^{-}(0) 
+ \Theta_{\gamma}^{+} (0) + \Theta_{\gamma}^{-}(0) \right]} 
 \right\rangle 
\right. \nonumber \\
&& \hspace{1.2in} + 
\left.
\left\langle 
e^{\frac{i}{\sqrt{2}}  
\left[ \Phi_{\gamma}^{+}(\tilde{r}) -\Phi_{\gamma}^{-}(\tilde{r}) 
- \Theta_{\gamma}^{+} (\tilde{r}) - \Theta_{\gamma}^{-} (\tilde{r}) \right]}
e^{-\frac{i}{\sqrt{2}}  
\left[ \Phi_{\gamma}^{+}(0) - \Phi_{\gamma}^{-}(0) 
- \Theta_{\gamma}^{+} (0) - \Theta_{\gamma}^{-}(0) \right]}
\right\rangle 
\right\}, 
\end{eqnarray}
\end{widetext}
where we used the new bosonic fields, $\Phi_{\gamma}^{\pm}$ and $\Theta_{\gamma}^{\pm}$, and defined $\tilde{r}\equiv(r,t)$.

The relevant cosine term in Eq.~(\ref{Eq:H_fb}) tends to order the $\Phi_{\gamma}^{+}$ field, which will be locked into one of the minima or maxima of the cosine, depending on the sign of $B_{Ov} \propto A_{0}$. The canonically conjugated $\Theta_{\gamma}^{+}$ field, on the other hand, will be disordered. Consequently, the correlation functions of the $\Phi_{\gamma}^{+}$ field will be constants, and those containing the $\Theta_{\gamma}^{+}$ field will be exponentially suppressed.~\cite{Giamarchi:2003,Meng:2013a,Meng:2013,Meng:2014} 
In addition, the local extrema of the RKKY interaction (Table~\ref{Tab:scattering}) are also exponentially suppressed in the presence of the feedback. The physical picture is that since the $(\Phi_{\gamma}^{+},\Theta_{\gamma}^{+})$ sector is gapped due to the sine-Gordon term, its contribution to the RKKY interaction is much less than the gapless $(\Phi_{\gamma}^{-},\Theta_{\gamma}^{-})$ sector. As a result, we may calculate the transverse spin susceptibility by simply neglecting the $\Phi_{\gamma}^{+}$ and $\Theta_{\gamma}^{+}$ fields~\cite{Braunecker:2009b} and following the same steps as in Sec.~\ref{SubSec:chi}. In comparison with Eq.~(\ref{Eq:chi}), the velocities are replaced by $\tilde{u}$, and $\tilde{\chi}_{AA}^{>,x}$ acquires an extra factor of $\frac{1}{2}$ because its first term in Eq.~(\ref{Eq:chi_greater_fb}) contains only the gapped $\Phi_{\gamma}^{+}$ field and is thus suppressed. In addition, the essential modification, namely the modified exponents, is 
\begin{eqnarray}
\tilde{g}_{x} = \tilde{g}_{y} &=& \frac{\tilde{K}}{2}.
\end{eqnarray}
As a result, the modified static spin susceptibility for $x,y$ components is given by
\begin{eqnarray}
\tilde{\chi}_{AA}^{x}(q) &=& \tilde{\chi}_{AA}^{y}(q) \nonumber\\
&=& - \frac{ \sin (\pi \tilde{g}_{x}) }{ 32 \pi^2 \hbar \tilde{u}}
\left(\frac{\tilde{\lambda}_{T}}{2\pi a} \right)^{2-2\tilde{g}_{x}} \nonumber\\
&&\times \sum_{\kappa=\pm}
\left| 
\frac{ \Gamma\left( 1-\tilde{g}_{x}\right) \Gamma\left( \frac{\tilde{g}_{x}}{2}-i\frac{\tilde{\lambda}_{T}}{4\pi}\left( 
q - 2 \kappa k_{F} \right) \right)}
{\Gamma\left(
\frac{2-\tilde{g}_{x}}{2}-i\frac{\tilde{\lambda}_{T}}{4\pi}\left( q - 2 \kappa k_{F} \right) \right)}
\right|^{2}, \nonumber\\
\label{Eq:chi_feedback}
\end{eqnarray}
where the thermal length now becomes $\tilde{\lambda}_{T}=\frac{\hbar \tilde{u}}{k_{B}T}$.

On the other hand, the $z$ component of the spin susceptibility is exponentially suppressed by the helical order gap and is much smaller than the transverse component. The full expression for $\tilde{\chi}_{\alpha\beta}^{z}(q)$ is difficult to compute because it involves the gapped (non-free) bosons.~\cite{Voit:1998,Starykh:1999} However, since the transition temperature is determined by $\tilde{\chi}_{\alpha\beta}^{x}(q)$ instead of $\tilde{\chi}_{\alpha\beta}^{z}(q)$, the full expression for $\tilde{\chi}_{\alpha\beta}^{z}(q)$ is not necessary. Nevertheless, to understand how the RKKY interaction depends on the gap, we compute the RKKY peak value at zero temperature~\cite{Meng:2013,Meng:2014} 
\begin{eqnarray}
\tilde{\chi}_{AA}^{z}(q=2k_{F})
&=& - \frac{ 1 }{4\pi \hbar \tilde{u}} \frac{1}{2-2 \tilde{g}_{z}} \left[ \left( \frac{\tilde{\Delta}_{a}}{\Delta_{m}} \right)^{(2-2 \tilde{g}_{z})} -1 \right], \nonumber\\
\label{Eq:chi_feedback_z}
\end{eqnarray}
with $\Delta_{m}$ being the gap due to the antiferromagnetic helix,  defined in Sec.~\ref{SubSec:RG}, and the modified exponent is
\begin{eqnarray}
\tilde{g}_{z} &=& \frac{1}{8} \left(\tilde{K} + \frac{1}{\tilde{K}}\right).
\end{eqnarray}
For CNTs, we obtain $\tilde{g}_{x} \approx 0.19$, and $\tilde{g}_{z} \approx 0.38$. More details about the calculation of the RKKY peak in gapped systems will be given in Sec.~\ref{SubSec:chi_gapped}, where the pairing gap due to the proximity effect is taken into account. We conclude that the anisotropic spin susceptibility due to the ordered spins serves an indirect experimental signature for the nuclear spin order.~\cite{Meng:2013}

\begin{figure}[htb]
\centering
\includegraphics[width=\linewidth]{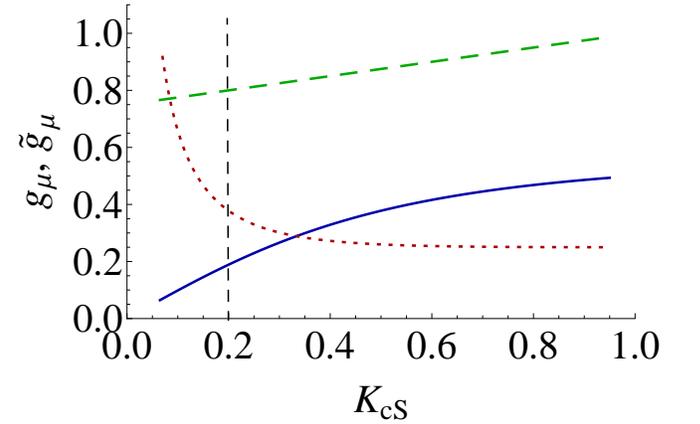}
\caption{Exponents $g_{\mu}$ and $\tilde{g}_{\mu}$ as a function of the Luttinger liquid parameter, $K_{cS}$. The blue solid and red dotted lines give the modified exponents $\tilde{g}_{x}=\tilde{g}_{y}$ and $\tilde{g}_{z}$ in the presence of the feedback, respectively. The green dashed curve describes the exponent $g_{\mu}$ without the feedback. The vertical black dashed line marks the value we use to evaluate, $K_{cS}=0.2$. The other parameters used here are the same as in Fig.~\ref{Fig:RKKY}. }
\label{Fig:exponents}
\end{figure}

\begin{figure}[htb]
\centering
\includegraphics[width=\linewidth]{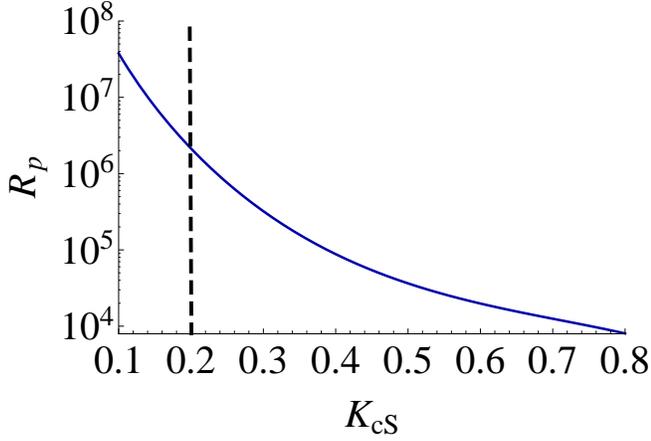}
\caption{Ratio of the RKKY peak values, $R_{p} \equiv |\tilde{J}_{\alpha\beta}^{x}(q=2k_{F})/J_{\alpha\beta}^{x}(q=2k_{F})|$ at  $T=50$~mK as a function of the Luttinger liquid parameter, $K_{cS}$. The other parameters are the same as in Fig.~\ref{Fig:RKKY}. The black dashed line marks the value $K_{cS}=0.2$.}
\label{Fig:peaks}
\end{figure}

The feedback-modified RKKY interaction is given by $\tilde{J}_{\alpha\beta}^{\mu}(q)=A_{0}^2a\tilde{\chi}_{\alpha\beta}^{\mu}(q)/2$.  The value of the RKKY peak mainly depends on the exponent, which depends strongly on the parameter $K_{cS}$, as demonstrated in Fig.~\ref{Fig:exponents}. The $z$ component of the static spin susceptibility is exponentially suppressed by the helical order gap, what results in a larger RKKY interaction in the transverse direction $|\tilde{J}_{\alpha\beta}^{x}(q)|>|\tilde{J}_{\alpha\beta}^{z}(q)|$, further stabilizing the planar magnetic order. Accordingly, this easy-plane anisotropy, in contrast to Ref.~\citenum{Braunecker:2009a,Braunecker:2009b}, naturally justifies the ansatz of the planar nuclear spin order. 
In Fig.~\ref{Fig:peaks} we plot the ratio of the RKKY peak value with the feedback to the one without the feedback as a function of the Luttinger liquid parameter, $K_{cS}$. The RKKY peak is strongly enhanced in the presence of the feedback, and the ratio increases with smaller $K_{cS}$, corresponding to stronger interaction. 

\begin{figure}[htb]
\centering
\includegraphics[width=\linewidth]{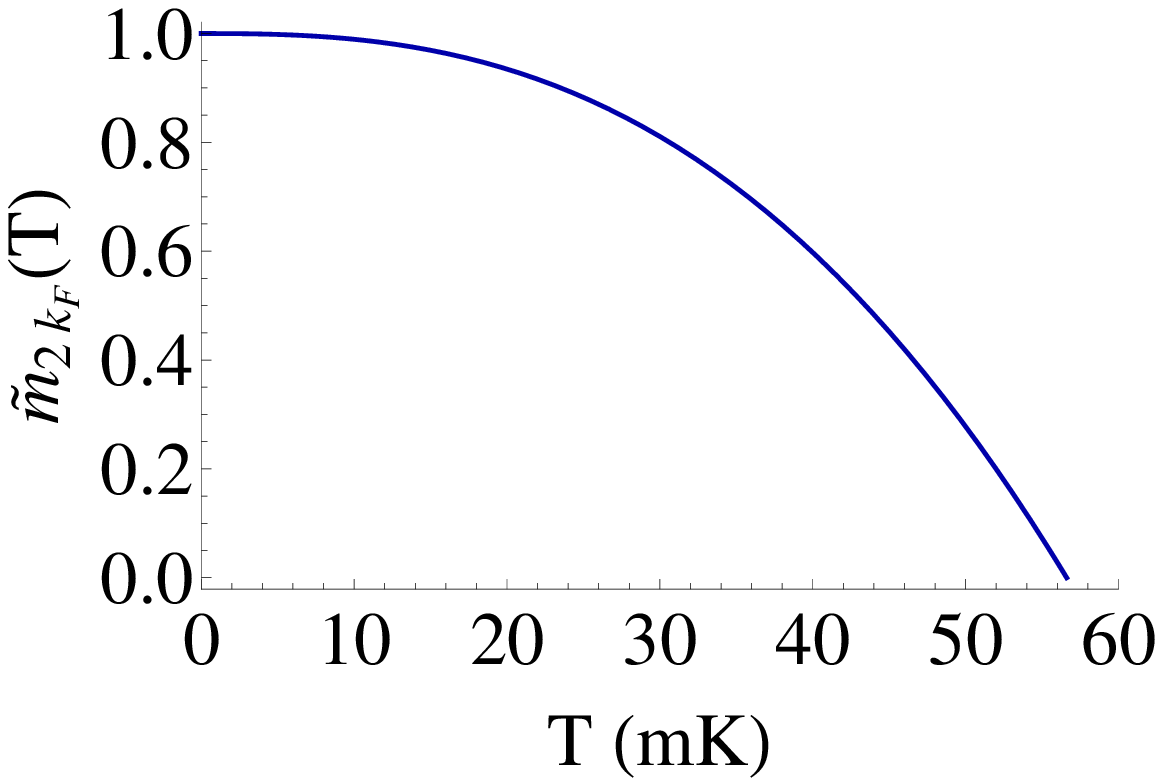}
\includegraphics[width=\linewidth]{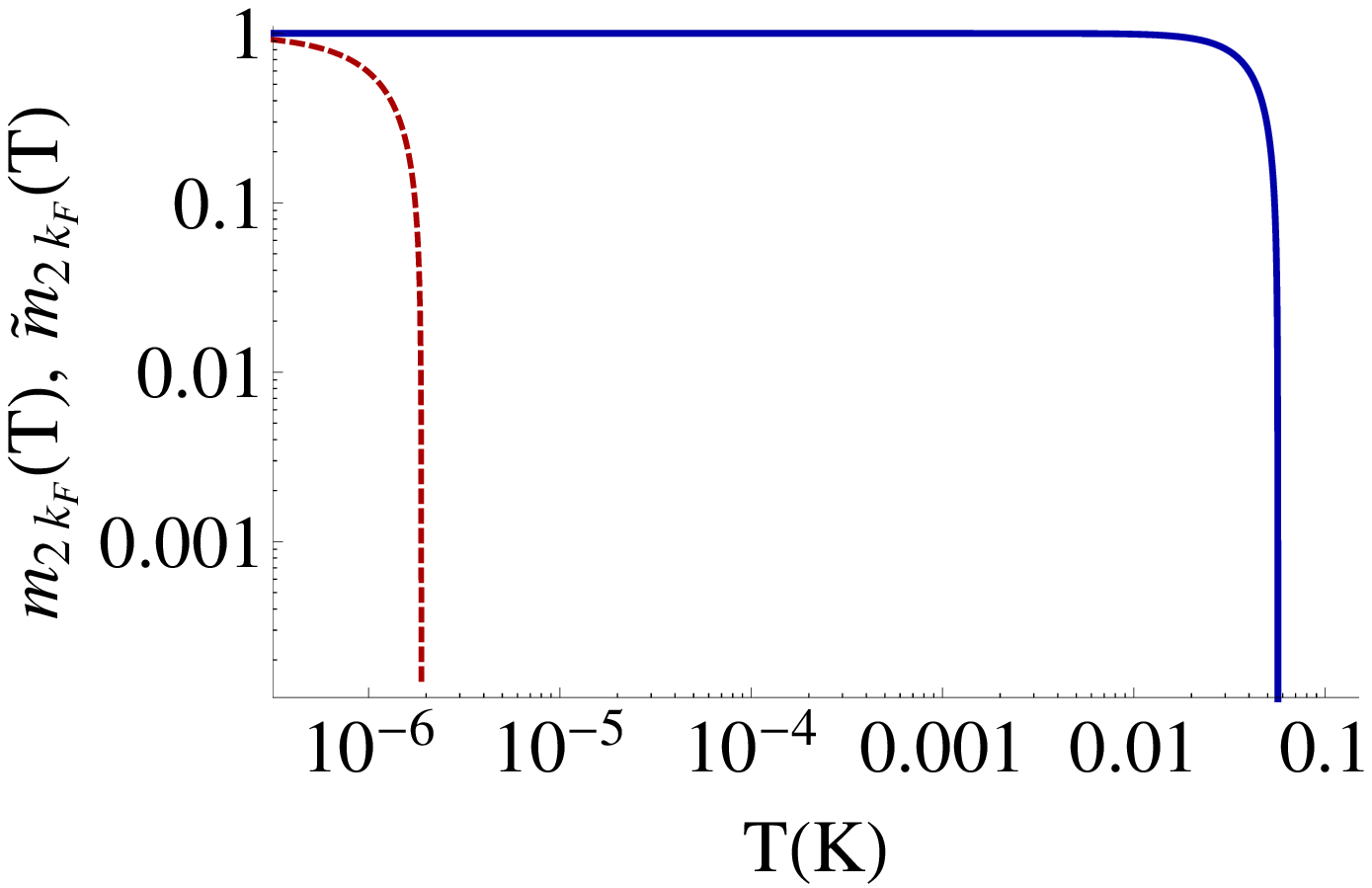}
\caption{Temperature dependence of the order parameter. Top: the order parameter in the presence of the feedback, $\tilde{m}_{2k_{F}}(T)$, from Eq.~(\ref{Eq:BlochLaw_fb}). Bottom: the order parameter without (red dashed) and with (blue solid) the feedback from Eqs.~(\ref{Eq:BlochLaw}) and (\ref{Eq:BlochLaw_fb}), respectively, in the logarithmic scale. The parameters used here are the same as in Fig.~\ref{Fig:RKKY}.}
\label{Fig:m2kF_fb}
\end{figure}

We evaluate the magnon spectrum using the modified RKKY interaction, and estimate the transition temperature, repeating the procedure described in Sec.~\ref{SubSec:Tc}. The temperature dependence of the order parameter is the same generalized Bloch law, 
\begin{eqnarray}
\tilde{m}_{2k_{F}}(T)
&=& 1 - \left( \frac{T}{\tilde{T}_{0}} \right)^{3-2\tilde{g}_{x}},
\label{Eq:BlochLaw_fb}
\end{eqnarray}
with a modified exponent, $(3-2\tilde{g}_{x})$, and a modified transition temperature, 
\begin{eqnarray}
k_{B}\tilde{T}_{0}
&\approx& \left[ \frac{I^2  A_{0}^2}{2N_{\perp}}
\left( \tilde{\Delta}_{a} \right)^{1-2 \tilde{g}_{x}} 
C(\tilde{g}_{x}) \right]^{\frac{1}{3-2\tilde{g}_{x}}},
\label{Eq:T0_fb}
\end{eqnarray}
which gives $\tilde{T}_{0} \approx 57$~mK as shown in Fig.~\ref{Fig:m2kF_fb}.
In comparison with the absence of the feedback, the transition temperature is enhanced by more than four orders of magnitude.

The feedback not only modifies the exponents and therefore strongly enhances the transition temperature, but also gaps out the $(\Phi_{\gamma}^{+},\Theta_{\gamma}^{+})$ modes. This leaves an effective Hamiltonian for the gapless $(\Phi_{\gamma}^{-},\Theta_{\gamma}^{-})$ modes, which mix the charge and spin sectors of the bosonic fields [see Eqs.~(\ref{Eq:newfield1}) and (\ref{Eq:newfield2})].
Consequently this produces a density-wave order that combines charge and spin degrees of freedom and reconstructs the electronic states.~\cite{Braunecker:2009a,Braunecker:2009b,Braunecker:2010} The combination of the charge and spin degrees of freedom signifies no spin-charge separation in this unusual Luttinger liquid, which is equivalent to introducing a synthetic spin-orbit interaction.~\cite{Giamarchi:2003}
Indeed, it has been shown that, upon a spin-dependent gauge transformation, a helical magnetic order is equivalent to spin-orbit interaction combined with Zeeman field.~\cite{Braunecker:2010,Klinovaja:2012a,Vazifeh:2013} This can also be seen from the form of the Overhauser field, a spatially oscillating field which combines the spin and orbital degrees of freedom. Since the spin-orbit interaction is crucial for non-trivial topology, we now consider coupling the system to a superconductor and discuss the realization of MFs in CNTs.

\section{RKKY interaction in the presence of superconductivity \label{Sec:RKKY_gapped}}
\subsection{Spin susceptibility in the presence of the pairing gap \label{SubSec:chi_gapped}}

Since the RKKY interaction in metallic phases is mediated by conduction electrons, it is a bit surprising that even in gapped phases there can still be nonvanishing RKKY peaks which give rise to nuclear spin orders.~\cite{Klinovaja:2013b,Meng:2013,Braunecker:2013} 
In this section we show that in the presence of the superconductivity the RKKY interaction can still form $q=\pm 2k_{F}$ peaks, even though the strength of the peaks are reduced by the pairing gap. 

Here we consider only BCS-type Cooper pairs with zero momentum. 
Since there are four Fermi points in metallic CNTs, Cooper pairs with zero momentum can be formed either between right movers at the $K_{+}$ valley and left movers at the $K_{-}$ valley, denoted as the exterior branches, or between left movers at the $K_{+}$ valley and right movers at the $K_{-}$ valley, denoted as the interior branches. 
In the presence of the proximity-induced superconductivity,~\cite{Kasumov:1999,Morpurgo:1999,Jarillo-Herrero:2006,Kleine:2009,Shimizu:2011} the pairing gaps for the exterior and interior branches are in general different.~\footnote[5]{ 
When a CNT is brought in contact with a superconductor, Cooper pairs from the superconductor can tunnel into the CNT through two processes. In one process a Cooper pair tunnels to either one of the sublattices with the pairing amplitude, $\Delta_{s,d}$, while in the other process the pairings are between electrons on different sublattices with the pairing amplitude, $\Delta_{s,n}$.
The pairing gaps of the exterior and interior branches are related to $\Delta_{s,d}$ and $\Delta_{s,n}$ by~\cite{Klinovaja:2012c}
\begin{equation*}
\Delta_{s}^{(e/i)} = \Delta_{s,d} \pm \Delta_{s,n} \left| \frac{\hbar v_{F} k}{ \sqrt{\alpha^2+(\hbar v_{F} k)^2} } \right|,
\end{equation*} 
where the curvature-induced $\alpha=-0.08$~meV$/R$[nm] with $R$ being the radius of the CNT.~\cite{Klinovaja:2011a,Klinovaja:2011b} 
Strictly speaking the pairing gap then depends on the wave vector. However, since in this work we are investigating the topological effect of the synthetic spin-orbit interaction rather than the intrinsic one, we drop $\alpha$, and
thus neglect the wave vector dependence of the pairing gaps. (See also Ref.~98 in Ref.~\citenum{Braunecker:2009b}.) Therefore, we have
$\Delta_{s}^{(e/i)} = \Delta_{s,d} \pm \Delta_{s,n}$, which are constants and may be considered as phenomenological parameters. }

Similar to Sec.~\ref{Sec:RKKY}, we shall first consider the system in the absence of feedback, and include the feedback afterward. The Hamiltonian now consists of the interacting conduction electron terms $\mathcal{H}_{\textrm{el}}$, given by Eq.~(\ref{Eq:LL}), and the pairing terms
\begin{eqnarray}
\mathcal{H}_{\textrm{s}} 
&=& g_{s} \sum_{\gamma,\sigma} \int dr \left[
\psi_{R,\gamma,\sigma}^{\dagger} (r) \psi_{L,\bar{\gamma},\bar{\sigma}}^{\dagger} (r) + h.c. \right] \nonumber\\
 &=& \frac{ g_{s} }{\pi a} \sum_{\gamma,\sigma} \int dr 
\cos \left( \gamma \phi_{cA} + \sigma \phi_{sS} - \theta_{cS} -\sigma \gamma \theta_{sA} \right), \nonumber\\
\end{eqnarray}
where $g_{s}$ is the coupling of the pairing terms, and in the second line the pairing terms are expressed in terms of the bosonic fields. 
We see that the pairing terms contain the $\phi_{sS}$ and $\theta_{cS}$ fields, whereas the feedback terms contain their conjugate fields, $\theta_{sS}$ and $\phi_{cS}$ [see Eq.~(\ref{Eq:H_s-G})]. Therefore, the antiferromagnetic nuclear spin helix and superconductivity compete with each other. The repulsive electron-electron interaction increases the feedback but reduces the pairing gap.~\cite{Gangadharaiah:2011,Stoudenmire:2011,Braunecker:2013} Nonetheless, we will see in Sec.~\ref{SubSec:refermionization} that the pairing terms are still RG relevant. 

The details of the calculation on the spin susceptibility in the presence of the superconductivity are given in Appendix~\ref{Sec:chi_gapped}. The zero-temperature value of the peaks of the static spin susceptibility reads 
\begin{eqnarray}
\chi_{AA}^{\mu}(q=2k_{F})
&\approx& -\frac{1}{4\pi \hbar v_{F}} \frac{1}{2-2g_{\mu}} 
\left[ \left( \frac{\Delta_{a}}{\Delta_{s}} \right)^{(2-2g_{\mu})} -1\right].\nonumber \\
\label{Eq:chi_gapped}
\end{eqnarray} 
For the noninteracting limit, $g_{\mu} \rightarrow 1$, we have
\begin{eqnarray}
\chi_{AA}^{\mu}(q=2k_{F})
\approx -\frac{1}{4\pi \hbar v_{F}} \ln \left( \frac{\Delta_{a}}{\Delta_{s}} \right), 
\end{eqnarray}
which recovers the logarithmic dependence of the RKKY susceptibility peak in noninteracting systems without valley degrees of freedom.~\cite{Klinovaja:2013b}

\begin{figure}[htb]
\centering
\includegraphics[width=\linewidth]{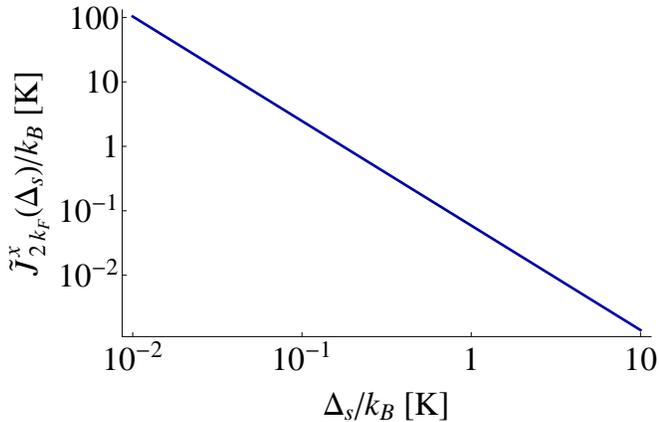}
\caption{RKKY peak value at zero temperature as a function of the pairing gap in the presence of the feedback. The parameters used here are the same as in Fig.~\ref{Fig:RKKY}.}
\label{Fig:RKKYfb_gapped}
\end{figure}

\subsection{Transition temperature in the presence of the pairing gap\label{SubSec:T_gapped}}

The peak value of the RKKY interaction in the presence of the pairing gap is given by
$J_{q=2k_{F}}^{\mu}(\Delta_{s}) = \frac{A_{0}^{2}a}{2} |\chi_{AA}^{\mu}(q=2k_{F})|$. In the presence of the antiferromagnetic helix, the feedback essentially modifies the exponents, $g_{\mu} \rightarrow \tilde{g}_{\mu}$. Therefore, following the same procedure, the enhanced RKKY interaction due to the feedback can be obtained as,
\begin{eqnarray}
\tilde{J}_{q=2k_{F}}^{x}(\Delta_{s}) \approx \frac{A_{0}^{2}}{16\pi \tilde{\Delta}_{a}} \frac{1}{2-2\tilde{g}_{x}} \left[ \left( \frac{\tilde{\Delta}_{a}}{\Delta_{s}} \right)^{(2-2\tilde{g}_{x})} -1 \right], \nonumber\\
\label{Eq:RKKYfb_gapped}
\end{eqnarray} 
where the modified exponent $\tilde{g}_{x}\approx 0.19$ was obtained in Sec.~\ref{SubSec:Tc_fb}. From Eq.~({\ref{Eq:RKKYfb_gapped}}) and Fig.~\ref{Fig:RKKYfb_gapped} we can see that the RKKY interaction depends on the pairing gap, or, more precisely, the ratio of the pairing gap to the bandwidth, $\Delta_{s}/\tilde{\Delta}_{a}$.

Assuming the induced superconducting gap is $\Delta_{s} = 0.2~$K, the peak value of the RKKY interaction is $\tilde{J}_{q=2k_{F}}^{x} \approx 0.8~$K. This relatively large peak value is due to the small ratio of the pairing gap to the bandwidth and small exponent $\tilde{g}_{x}$ in CNTs. However, as can be seen in Fig.~\ref{Fig:RKKYfb_gapped}, the RKKY peak drops quickly with an increasing pairing gap. Using Eq.~(\ref{Eq:m2kF}), $\hbar \omega_{m} \rightarrow 2I \tilde{J}_{q=2k_{F}}^{x} /N_{\perp}$, and Eq.~\eqref{Eq:RKKYfb_gapped}, we obtain the transition temperature of the antiferromagnetic helix, 
\begin{equation}
k_{B} \tilde{T}_{0} \approx \frac{I^2}{N_{\perp}} \tilde{J}_{q=2k_{F}}^{x} , 
\label{Eq:T0_gapped}
\end{equation}
reduced by the induced pairing gap to $k_{B} \tilde{T}_{0} \approx 17~\textrm{mK}$. We note that in contrast to Eq.~(\ref{Eq:T0_fb}), the transition temperature $\tilde{T}_{0}$ now scales as $N_{\perp}^{-1}$ instead of $N_{\perp}^{-1/(3-2\tilde{g}_{x})}$. We also note that the transition temperature given in Eq.~(\ref{Eq:T0_gapped}) is overestimated because the temperature dependence of $\tilde{J}_{q=2k_{F}}^{x}$ is not included in Eq.~\eqref{Eq:RKKYfb_gapped}. However, this overestimate is negligible for low temperature, $k_{B}T \ll \Delta_{s}$.
 
In the previous estimation, we assume the pairing gap of $\Delta_{s} = 0.2~$K, somewhat smaller than a typical pairing gap ($\sim$1K) of parent superconductors used for the proximity effect.  
Such reduction is expected due to the electron-electron interactions,~\cite{Gangadharaiah:2011,Stoudenmire:2011} and will be further discussed in Sec.~\ref{SubSec:refermionization}. Alternatively, one may intentionally reduce the induced gap; for example, inserting a graphene sheet between a CNT and a superconductor. On the other hand, while a smaller $\Delta_{s}$ is beneficial to a higher $\tilde{T}_{0}$, it also results in a longer localization length for Majorana fermions (MFs), as discussed in Sec.~\ref{Sec:discussion}, so that a trade-off between the two parameters needs to be considered.

\section{Topological superconductivity \label{Sec:Top_Su}}
\subsection{Refermionization \label{SubSec:refermionization}}

We now consider the possibility to realize MFs in our system. In Refs.~\onlinecite{Gangadharaiah:2011,Stoudenmire:2011} it was shown that MFs may survive even in the presence of very strong electron-electron interactions, if the pairing term is RG relevant (in other words, if the interactions do not eliminate the pairing gap). In this case, the interacting bosonic Hamiltonian with spin-orbit interaction, Zeeman field, and pairing terms can be mapped onto a {\it noninteracting} fermion model with a reduced pairing gap. We therefore first establish that this is the case here, too. 

To this end, we write the pairing terms in the $(\Phi_{\gamma}^{\pm}, \Theta_{\gamma}^{\pm})$ basis introduced in Sec.~\ref{SubSec:RG}, 
\begin{eqnarray}
\mathcal{H}_{\textrm{s}} 
&=& g_{s} \sum_{\gamma,\sigma} \int dr \left[
\psi_{R,\gamma,\sigma}^{\dagger} (r) \psi_{L,\bar{\gamma},\bar{\sigma}}^{\dagger} (r) + h.c. \right] \nonumber\\
 &=& \frac{ g_{s} }{\pi a} \int dr \left[
\cos \left( -\Phi_{+}^{-} + \Phi_{-}^{-} + \Theta_{+}^{-} + \Theta_{-}^{-} \right) 
\right. \nonumber\\
&& \hspace{0.2in} \left. 
+ \cos \left( \Phi_{+}^{-} - \Phi_{-}^{-} + \Theta_{+}^{-} + \Theta_{-}^{-} \right) \right],
\label{Eq:LL_pairing}
\end{eqnarray} 
which scales as
\begin{eqnarray}
\left\langle \cos \left( \mp \Phi_{+}^{-} \pm \Phi_{-}^{-} + \Theta_{+}^{-} + \Theta_{-}^{-} \right) \right\rangle \propto \left( \frac{a}{r} \right)^{\frac{1}{2}\left( \tilde{K} + \frac{1}{\tilde{K}} \right)}, \nonumber\\
\end{eqnarray}
with $\tilde{K}$ given in Eq.~\eqref{Eq:k_tilde}.
Since the pairing gaps for exterior and interior branches have the same scaling dimensions, we do not distinguish them in Eq.~\eqref{Eq:LL_pairing} to simplify the RG analysis.
Thus, the RG flow equation for the dimensionless coupling $\tilde{y}_{s}(l)\equiv g_{s}(l)/\tilde{\Delta}_{a}(l)$ reads~\cite{Giamarchi:2003}
\begin{eqnarray}
\frac{d \tilde{y}_{s}(l)}{d l} = \left[2 -\frac{1}{2}\left( \tilde{K} + \frac{1}{\tilde{K}} \right)\right] \tilde{y}_{s}(l),
\end{eqnarray}
which gives the condition for the pairing term to be RG relevant, 
\begin{eqnarray}
2-\sqrt{3} < \tilde{K} < 2+\sqrt{3}. 
\end{eqnarray}
For our parameters, we have $\tilde{K} \approx 0.38$, so the pairing term is relevant. This also justifies the reduced gap used to determine the order of magnitudes of the RKKY peak in Sec.~\ref{SubSec:T_gapped}. 

Here we briefly summarize the operators involved in the procedure. The feedback (Overhauser field), which gaps out the $(L,\uparrow)$ and $(R,\downarrow)$ particles within each valley, can be described as $\cos\left( 2 \Phi_{\gamma}^{+}\right)$ with the scaling dimension $\tilde{K}$. On the other hand, the pairing terms are written in terms of $\cos \left( \mp \Phi_{+}^{-} \pm \Phi_{-}^{-} + \Theta_{+}^{-} + \Theta_{-}^{-} \right)$ with the scaling dimension $( \tilde{K} + 1/\tilde{K} )/2$. While these two terms compete with each other, as discussed in Sec.~\ref{SubSec:chi_gapped}, both of them are relevant for the parameters of CNTs. 

We now consider distinct exterior and interior pairing gaps,~\cite{LeHur:2008,Klinovaja:2012c} defined as $\Delta_{s}^{(e)}$ and $\Delta_{s}^{(i)}$, respectively. We then recast the Hamiltonian into a noninteracting fermionic model through the refermionization procedure.
To be explicit, we define the slowly varying fields $R_{\gamma,\sigma}$ and $L_{\gamma,\sigma}$ such that
$\psi_{\gamma,\sigma} (r) = R_{\gamma,\sigma}e^{ik_{F}r} + L_{\gamma,\sigma}e^{-ik_{F}r}$, and they are related to the bosonic fields, $\phi_{\gamma\sigma}$ and $\theta_{\gamma\sigma}$, defined in Eq.~(\ref{Eq:bosonization}), by
\begin{subequations}
\label{Eq:refermionization}
\begin{eqnarray}
R_{\gamma,\sigma} &\equiv& \frac{1}{\sqrt{2\pi a}} e^{i \left[-  \phi_{\gamma\sigma}(r) + \theta_{\gamma\sigma}(r) \right]}, \\
L_{\gamma,\sigma} &\equiv& \frac{1}{\sqrt{2\pi a}} e^{i \left[  \phi_{\gamma\sigma} (r) + \theta_{\gamma\sigma}(r) \right]}.
\end{eqnarray}
\end{subequations}
After the transformation, we obtain
\begin{eqnarray}
\mathcal{H}_{\textrm{top}} &=& \frac{1}{2} \int dr \phi^{\dagger}(r) \left[ -i \hbar v_{F} \tau_{3} \partial_{r} + \frac{\Delta_m}{2} \eta_{3}\left( \sigma_{1}\tau_{1} + \sigma_{2}\tau_{2} \right) \right. \nonumber\\
&& \hspace{0.85in}
+ \Delta_{s,+} \eta_{2}\delta_{1}\sigma_{2}\tau_{1}
-\Delta_{s,-} \eta_{2}\delta_{2}\sigma_{2}\tau_{2}  \nonumber\\
&& \hspace{0.85in} \left.
+\Delta_{Z} \eta_{3}\sigma_{3}
\right] \phi(r),
\label{Eq:H_top_r} 
\end{eqnarray}
where the 16-component spinor $\phi^{\dagger}(r)$ is defined as
\begin{eqnarray}
\phi^{\dagger}(r) &\equiv& \left( R_{+,\uparrow}^{\dagger},L_{+,\uparrow}^{\dagger},
R_{+,\downarrow}^{\dagger},L_{+,\downarrow}^{\dagger},
R_{-,\uparrow}^{\dagger},L_{-,\uparrow}^{\dagger},
R_{-,\downarrow}^{\dagger},L_{-,\downarrow}^{\dagger}, \right. \nonumber\\
&& \left.
R_{+,\uparrow},L_{+,\uparrow}, R_{+,\downarrow},L_{+,\downarrow},
R_{-,\uparrow},L_{-,\uparrow}, R_{-,\downarrow},L_{-,\downarrow} \right). \nonumber\\
\end{eqnarray}
Further, $\eta_{\mu}$, $\delta_{\mu}$, $\sigma_{\mu}$, and $\tau_{\mu}$ are Pauli matrices acting on particle-hole, valley, spin, and right/left degrees of freedom, respectively.
$\Delta_{m}$ is the gap due to the antiferromagnetic helix defined in Sec.~\ref{SubSec:RG} and $\Delta_{s,\pm}$ are defined as
\begin{eqnarray}
\Delta_{s,\pm} &\equiv& \frac{\Delta_{s}^{(e)} \pm  \Delta_{s}^{(i)}}{2}.
\end{eqnarray}
To find details how $\Delta_{s}^{(e/i)}$ evolves with the interaction would require the full RG analysis. Instead, we guide ourselves by Refs.~\citenum{Gangadharaiah:2011,Stoudenmire:2011} and estimate that the gap is reduced to one order smaller than that of the parent superconductor.
Since the typical gap of the parent superconductor used for proximity effect is of order kelvin,~\cite{Kasumov:1999,Morpurgo:1999,Jarillo-Herrero:2006,Kleine:2009,Shimizu:2011}  $\Delta_{s}^{(e)} = \Delta_{s}^{(i)}=0.2~$K are taken in the previous sections for the purpose of estimation. From now on we shall keep $\Delta_{s}^{(e/i)}$ to be
unfixed parameters, and hence $\Delta_{s,-}$ is nonzero in general. 
Finally, in Eq.~\eqref{Eq:H_top_r} we also included the Zeeman term, $\Delta_{Z}$, arising from a magnetic field perpendicular to the helical plane (along the tube). We do this to break the time-reversal symmetry,~\footnote[6]{While the physical time-reversal symmetry is broken by the nuclear spin helix, there exists (pseudo-)time-reversal symmetry, corresponding to a combination of flipping spins and interchanging the sublattice sites in the absence of the external magnetic fields.}
which has in general profound effects on MFs. 
Even though the magnetic field along the tube also induces orbital effects,~\cite{Klinovaja:2011a,Klinovaja:2011b} we have checked, by exact diagonalization, that adding them does not lead to any new gapped (topological) regime in the parameter space. Hence, we do not include such effects in Eq.~\eqref{Eq:H_top_r} for simplicity.

In momentum space, the bulk Hamiltonian is characterized by a matrix,
\begin{eqnarray}
H_{\textrm{top}}(k) &=& \hbar v_{F} k \, \tau_{3} + \frac{\Delta_m}{2} \eta_{3}\left( \sigma_{1}\tau_{1} + \sigma_{2}\tau_{2} \right) 
+ \Delta_{s,+} \eta_{2}\delta_{1}\sigma_{2}\tau_{1} \nonumber\\
&& 
-\Delta_{s,-} \eta_{2}\delta_{2}\sigma_{2}\tau_{2}
+\Delta_{Z} \eta_{3}\sigma_{3},
\label{Eq:H_top}
\end{eqnarray}
following from Eq.~\eqref{Eq:H_top_r} upon replacing $-i\partial_{r} \rightarrow k$, which allows us to inspect the symmetries of the Hamiltonian.~\footnote[7]{To understand the symmetry class of the Hamiltonian, $H_{\textrm{top}}(k)$, one usually considers the particle-hole and time-reversal symmetries of Eq.~\eqref{Eq:H_top}. 
The particle-hole symmetry is described by~\cite{Ryu:2010,Hasan:2010} 
\begin{equation*}
\Xi H_{\textrm{top}} (k) \Xi^{-1} = -H_{\textrm{top}}(-k),
\end{equation*}
where $\Xi=U_{P} \mathcal{K}$ with the unitary operator, $U_{P}$, and the complex conjugate, $\mathcal{K}$. However, there is ambiguity in the value of $\Xi^2$. For example, while both the choices of $U_{P}=\eta_{1}$ and $U_{P}=\eta_{2} \otimes \delta_{3}$ satisfy the above equation, they give different values of $\Xi^2$ and therefore different symmetry classes. In addition, a time-reversal invariant Hamiltonian satisfies~\cite{Ryu:2010,Hasan:2010} 
\begin{equation*}
\Theta H_{\textrm{top}} (k) \Theta^{-1} = H_{\textrm{top}}(-k),
\end{equation*}
with $\Theta=U_{T} \mathcal{K}$. Similarly, there is ambiguity in the value of $\Theta^2$ if $\Delta_{Z}=0$. For instance, the choice of $U_{T}=i \delta_{2} \otimes \sigma_{2} \otimes\tau_{2}$ gives $\Theta^2=-1$ whereas $U_{T}=\eta_{2} \otimes \delta_{1} \otimes \sigma_{2} \otimes\tau_{1}$ gives $\Theta^2=+1$. 
In order to avoid the ambiguity and investigate the topological properties of the Hamiltonian, we explicitly solve the problem, as discussed in the main text.
}
We note that while Eq.~(\ref{Eq:H_top}) describes a noninteracting model, it retains the features of Luttinger liquid through the renormalized gap parameters, $\Delta_{m}$ and $\Delta_{s}^{(e/i)}$.

\subsection{Topological superconductivity and MFs \label{SubSec:MF}}

The Hamiltonian is block diagonal if decomposed into two pieces, $\mathcal{H}_{\textrm{top}}=\mathcal{H}_{\textrm{top}}^{(1)}+\mathcal{H}_{\textrm{top}}^{(2)}$, with
\begin{eqnarray}
\mathcal{H}_{\textrm{top}}^{(j)} &=& \frac{1}{2} \int dr \; \phi_{j}^{\dagger}(r) H_{\textrm{top}}^{(j)}(r) \phi_{j}(r),
\label{Eq:H_top_block}
\end{eqnarray}
where the 8-component spinor $\phi_{1}^{\dagger}(r)$ is formed by the fields gapped by the nuclear spin helix, and $\phi_{2}^{\dagger}(r)$ is formed by the other fields, explicitly, 
\begin{eqnarray}
\phi_{1}^{\dagger}(r) &\equiv& 
(L_{+,\uparrow}^{\dagger}, R_{+,\downarrow}^{\dagger},
L_{-,\uparrow}^{\dagger}, R_{-,\downarrow}^{\dagger},
L_{+,\uparrow}, R_{+,\downarrow},
L_{-,\uparrow},R_{-,\downarrow}), \nonumber \\
\phi_{2}^{\dagger}(r) &\equiv& 
(R_{+,\uparrow}^{\dagger}, L_{+,\downarrow}^{\dagger}, 
R_{-,\uparrow}^{\dagger}, L_{-,\downarrow}^{\dagger}, 
R_{+,\uparrow}, L_{+,\downarrow},
R_{-,\uparrow}, L_{-,\downarrow}).  \nonumber \\
\end{eqnarray}
The corresponding 8-by-8 Hamiltonian densities $H^{(1)}_{\textrm{top}}(r)$ and $H^{(2)}_{\textrm{top}}(r)$ are obtained from Eq.~(\ref{Eq:H_top_r}). 

\begin{figure}[htb]
\centering
\includegraphics[width=\linewidth]{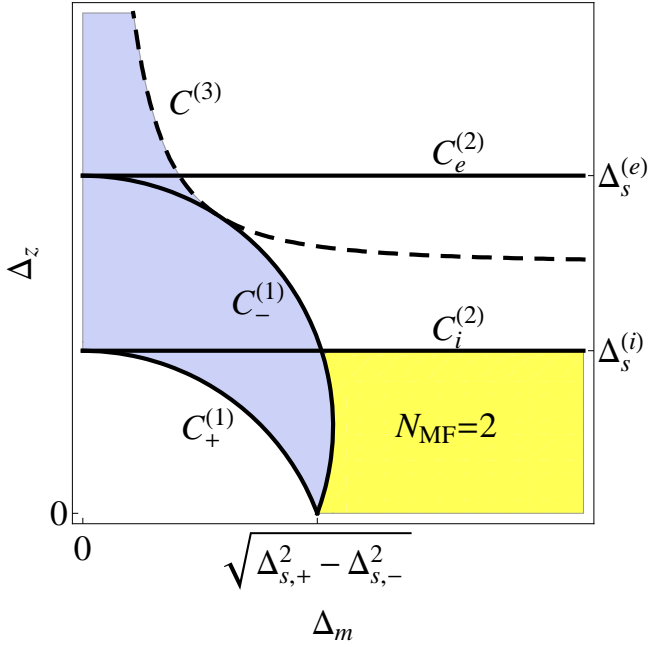}
\caption{Phase diagram on the $\Delta_{m}$-$\Delta_{Z}$ plane. The black solid curves are marked as $C_{+}^{(1)}$, $C_{-}^{(1)}$, $C_{e}^{(2)}$, and $C_{i}^{(2)}$, whereas the black dashed curves is marked as $C^{(3)}$. The intercepts of these curves on the axes are also labeled. The yellow shaded region corresponds to the MF regime. In the blue region the bulk spectrum is gapless. }
\label{Fig:phase}
\end{figure}

To find the MF solutions, we solve the Schr\"odinger equation at zero energy while imposing the self-conjugation and boundary conditions on the wave functions.~\cite{Klinovaja:2012a} The details of the calculation are given in Appendix~\ref{Sec:MF_solution}. The results of this procedure are summarized in the phase diagram shown in Fig.~\ref{Fig:phase}. We plot it in the first quadrant of the $\Delta_{m}$-$\Delta_{Z}$ plane, where the other three quadrants can be obtained by inversion symmetry about the $\Delta_{m}$ and $\Delta_{Z}$ axes. The formulas for the curves, $C_{\pm}^{(1)}$, $C_{e/i}^{(2)}$, and $C^{(3)}$, are given in Appendix~\ref{Sec:MF_solution}. 
Both $C_{+}^{(1)}$ and $C_{-}^{(1)}$ intersect with the $\Delta_{m}$ axis at $\Delta_{m}=\sqrt{\Delta_{s,+}^2-\Delta_{s,-}^2}$. $C_{e}^{(2)}$ and $C_{i}^{(2)}$ intersect with the $\Delta_{Z}$ axis at $\left| \Delta_{Z} \right| = \left| \Delta_{s}^{(e)} \right|= \left| \Delta_{s,+} + \Delta_{s,-}\right|$ and $\left|\Delta_{Z}\right|=\left|\Delta_{s}^{(i)}\right|=\left|\Delta_{s,+} - \Delta_{s,-}\right|$, respectively. The blue region corresponds to a nontopological gapless state. 
The yellow shaded region corresponds to a regime with two MFs at one given end of the nanotube, and is defined by the inequalities,
\begin{subequations}
\label{Eq:MFs}
\begin{eqnarray}
\left( \Delta_{Z} \pm \Delta_{s,-} \right)^2 + \Delta_{m}^2 - \Delta_{s,+}^2 &>&0, \\
\left| \Delta_{Z} \right| - \left| \Delta_{s}^{(e)} \right| &<& 0, \\
\left| \Delta_{Z} \right| - \left| \Delta_{s}^{(i)} \right| &<& 0.
\end{eqnarray}
\end{subequations}
From Fig.~\ref{Fig:phase} we can see that neither the distinct exterior and interior pairing gaps nor the Zeeman field is necessary for this MF regime. 
We emphasize that the gap parameters, $\Delta_{m}$ and $\Delta_{s}^{(e/i)}$, are modified by the electron-electron interaction, and therefore reflect the features of Luttinger liquid in Fig.~\ref{Fig:phase}. 

The MF wave functions have composite nature and display multiple decay length scales, resulting in oscillations in addition to the exponential decay. 
The localization length, $\xi_{\textrm{loc}}$, is determined by the largest length scale of the inverses of $\kappa_{1,\pm}$ and $\kappa_{2,e/i}$ 
(defined in Appendix~\ref{Sec:MF_solution}). If the system is deep inside the MF regime, then $\Delta_{m}$ is the largest energy scale of the parameters, and the localization length will be determined by the smaller of $\Delta_{s}^{(e)}$ and $\Delta_{s}^{(i)}$; namely, 
\begin{eqnarray}
\xi_{\textrm{loc}}&=& \left\{
\begin{array}{lr}
\left(\kappa_{2,e} \right)^{-1} =\frac{\hbar v_{F}}{ \sqrt{\left[\Delta_{s}^{(e)}\right]^2 -\Delta_{Z}^2} }, &\; \textrm{if } 
\left| \Delta_{s}^{(e)} \right| < \left| \Delta_{s}^{(i)} \right|, \\
\left(\kappa_{2,i}\right)^{-1} =\frac{\hbar v_{F}}{\sqrt{\left[\Delta_{s}^{(i)}\right]^2 -\Delta_{Z}^2}}, &\; \textrm{if } 
\left| \Delta_{s}^{(e)} \right| > \left| \Delta_{s}^{(i)} \right|.
\end{array}
\right. \nonumber\\
\end{eqnarray}
For zero magnetic field, the localization length is the inverse of the smaller of the pairing gaps.

In addition to the magnetic field along the tube, we have also examined that adding a magnetic field perpendicular to the tube, i.e. $\Delta_{Z} \eta_{3}\sigma_{1}$, does not lead to any new gapped regime in the parameter space and therefore does not generate topological phases with single MFs, either.
Utilizing the MF wave functions for $\Delta_{Z}=\Delta_{s,-}=0$, in which case the analytical solutions are available, we have checked that the MF pair is not mixed by a perturbation $\delta H$ by evaluating the matrix element $\langle \Phi_{\textrm{MF},1}| \delta H | \Phi_{\textrm{MF},2} \rangle$. We found it is zero for $\delta H$ corresponding to the Zeeman field perpendicular to the tube, the distinct pairing gaps, the Zeeman field along the tube (including the orbital effects), and an electrostatic impurity potential. The fact that the MF pair is not hybridized by any of these terms further confirms its robustness.
It is interesting to note that the MF pair that we find is not split in spite of the lifted degeneracy in the bulk spectrum due to the broken time-reversal symmetry by the external magnetic field.
Through the explicit calculation, we conclude that Eq.~\eqref{Eq:H_top} supports topological phases with multiple MFs. It is remarkable that the criterion for the MFs, Eq.~\eqref{Eq:MFs}, is fulfilled for the parameters of CNTs.

\section{Discussion \label{Sec:discussion}}

In the absence of experimental techniques with atomic resolution, direct detections of the locally antiferromagnetic nuclear spin helix are even more challenging than the ferromagnetic one, since the magnetization signals~\cite{Zhou:2010,Grinolds:2013} average out to zero due to the sign changes of the susceptibility between different sublattice sites.~\cite{Stano:2013}  
Indirect experimental signatures of the nuclear spin order, however, can be searched for below the transition temperature. As discussed in the literature, these include: (1) the reduction of conductance by a factor of 2 due to the opening of the partial gap;~\cite{Braunecker:2009a,Braunecker:2009b,Scheller:2014,Meng:2014b} (2) the anisotropic spin susceptibility $\tilde{\chi}_{\alpha\beta}^{x}(q) \neq \tilde{\chi}_{\alpha\beta}^{z}(q)$ due to the formation of the nuclear spin order;~\cite{Meng:2013} (3) NMR response at the frequency set by the RKKY exchange due to the singular RKKY peak;~\cite{Stano:2014} (4) the unusual temperature dependence of the nuclear spin relaxation rate due to the Luttinger liquid parameters modified by the Overhauser field;~\cite{Zyuzin:2014} (5) the reentrant behavior in the conductance as a function of gate voltage due to the nuclear spin induced gap;~\cite{Rainis:2014} (6) the dynamical nuclear polarization at zero external magnetic field.~\cite{Kornich:2015}

Furthermore, experimental probes can be implemented to observe the distinct pairing gaps, $\Delta_{s}^{(e/i)}$. In general, there should be double-gap features below the superconducting critical temperature,~\cite{LeHur:2008} and the gap values should be reduced by the electron-electron interaction.~\cite{Klinovaja:2012c} Similarly, the helical gap due to the Overhauser field, $\Delta_{m}$, can be observed below the transition temperature, $\tilde{T}_{0}$, which decreases in the presence of the pairing gap. Interestingly, it has been reported that the NMR measurement of the double-wall CNTs, consisting of 89\% $^{13}$C enriched inner walls and natural 1.1\% $^{13}$C outer walls, revealed the formation of a spin gap at low temperatures.~\cite{Singer:2005} In addition, since the remaining gapless modes have definite helicity, CNTs may thus serve as spin filters, similar to the proposal in Ref.~\citenum{Braunecker:2010}.

The localization length of MFs is set by the smaller of $\Delta_{s}^{(e)}$ and $\Delta_{s}^{(i)}$. 
For $\Delta_{s}^{(e/i)}=0.2\text{--}2$~K and $\Delta_{Z}=0$, we obtain $\xi_{\textrm{loc}} \approx 3 \text{--} 30~\mu$m, so nanotubes with length $L \apprge 3~\mu$m are needed to avoid the overlap between MFs from the two ends.~\footnote[8]{Throughout the main text we have used a conservative estimate of $A_{0}$, which is one order smaller than the measured value.~\cite{Churchill:2009a,Churchill:2009b} If we instead take the measured value, $A_{0} \approx 100~\mu$eV for our estimation, then the transition temperature is greatly enhanced to $\tilde{T}_{0} \approx 110~$mK even with a pairing gap of $\Delta_{s} = 2~$K, resulting in a much shorter localization length, $\xi_{\textrm{loc}} \approx 3~\mu$m. 
}  
While increasing $\Delta_{s}^{(e/i)}$ leads to a shorter $\xi_{\textrm{loc}}$, a larger $\Delta_{s}^{(e/i)}$ substantially suppresses the transition temperature for the nuclear spin order, so there is a trade-off between high $\tilde{T}_{0}$ and short localization length.

Recently, a realization of MFs in armchair CNTs driven by external electric fields has been proposed,~\cite{Klinovaja:2012c} where the electric fields induce the helical modes,~\cite{Klinovaja:2011a,Klinovaja:2011b} a necessary prerequisite for MFs. 
However, those electric-field-induced MFs require fine tuning of the chemical potential, in contrast to the RKKY systems in the present work. Here, since the antiferromagnetic nuclear spin helix, resulting from the scattering between right-moving down-spin and left-moving up-spin electrons, always opens a gap at the Fermi surface (Fig.~\ref{Fig:helix_scattering}), the RKKY system does not require experimentally challenging fine tuning the chemical potential.
In addition, with the RKKY mechanism it is unnecessary to apply an external magnetic field, which is detrimental to the parent superconductor. Further, our calculation applies to any conducting CNTs, and therefore does not rely on a particular chirality of CNTs.~\footnotemark[2]
In comparison with the recently proposed spin-orbit coupled wires,~\cite{Oreg:2010,Lutchyn:2010} $^{13}$C nanotubes also have the advantage to explore MFs, owing to the aforementioned self-tuning properties and the availability of high-quality samples.~\cite{Laird:2014,Sanchez-Valencia:2014}
On the other hand, since a large pairing gap reduces the RKKY interaction and therefore $\tilde{T}_{0}$, parent superconductors with suitable pairing gaps are necessary to obtain both sufficiently high $\tilde{T}_{0}$ and short $\xi_{\textrm{loc}}$. 

Finally, we remark that the RKKY mechanism discussed here should also apply to other quasi-one-dimensional bipartite materials, such as metallic graphene nanoribbons, in which hyperfine interaction is nonvanishing~\cite{Fischer:2009} and the conduction electrons mediate the RKKY interaction.~\cite{Klinovaja:2013a}
In addition to the isotopically enriched materials, the antiferromagnetic helix can in principle be realized using magnetically doped systems, where carbon atoms are substituted by magnetic atoms, or magnetic atoms are deposited on the material. The substitutional or top-adsorbed magnetic atoms provide localized spins associated with a single site,~\cite{Power:2013} which can take the role of the $^{13}$C atoms discussed in this work, so we expect that the RKKY interaction can induce an antiferromagnetic helix in such configurations. 
For the plaquette (center-adsorbed) or bridge adatoms, on the other hand, the magnetic adatoms interact with an equal number of different sublattice sites,~\cite{Power:2013} so the sublattice-dependent oscillating terms with $q=2k_{F}$ in the spin susceptibility cancel out, provided that the couplings between the magnetic adatoms and the conduction electrons on different sublattice sites are equal. 
As a result, we do not expect the antiferromagnetic helix to be realized in such configurations. However, the remaining {\it sublattice-independent} oscillating terms with $q=2(k_{v}\pm k_{F})$ can still lead to a ferromagnetic helical order,~\cite{Braunecker:2009a,Braunecker:2009b} where the RKKY peaks with different momenta result in a beating pattern, as in two-subband quantum wire systems.~\cite{Meng:2013a} We also note that in graphene at half filling, it was found that the plaquette or bridge adatoms lead to a cancellation of the oscillations in the RKKY interaction,~\cite{Saremi:2007,Brey:2007,Black-Schaffer:2010} and therefore no kind of helical order will be realized in this case.

\acknowledgments
We acknowledge support from the Swiss NSF and NCCR QSIT.

\appendix

\section{\label{Sec:chi_gapped} Spin susceptibility in the presence of superconductivity}

In this appendix we calculate the spin susceptibility in the presence of the pairing gap. As given in Sec.~\ref{SubSec:chi_gapped}, the pairing terms take the form,
\begin{eqnarray}
\mathcal{H}_{\textrm{s}} 
&=& \frac{ g_{s} }{\pi a} \sum_{\gamma,\sigma} \int dr 
\cos \left( \gamma \phi_{cA} + \sigma \phi_{sS} - \theta_{cS} -\sigma \gamma \theta_{sA} \right). \nonumber\\
\end{eqnarray}
Expanding the sine-Gordon term around its minimum and keeping only the second-order terms, we obtain~\cite{Giamarchi:2003}
\begin{eqnarray}
\mathcal{H}_{\textrm{s}} 
&\approx& \frac{ \Delta_{s}^2 }{2\pi \hbar v_{F}} \int dr 
 \left( \phi_{cA}^2 + \phi_{sS}^2 + \theta_{cS}^2 + \theta_{sA}^2 \right),
\end{eqnarray}
where $\Delta_{s} \equiv 2 \sqrt{g_{s} \Delta_{a}}$ is the proximity-induced pairing gap. 
In the presence of distinct exterior and interior pairing gaps, the pairing gap is replaced with
\begin{equation}
\Delta_{s} \rightarrow \sqrt{ \frac{ \left[\Delta_{s}^{(e)}\right]^2 + \left[\Delta_{s}^{(i)}\right]^2}{2} },
\end{equation}
where $\Delta_{s}^{(e)}$ and $\Delta_{s}^{(i)}$ are the pairing gaps for the exterior and interior branches, respectively. However, for simplicity we shall set $\Delta_{s}^{(e)}=\Delta_{s}^{(i)}=\Delta_{s}$ in this appendix. We note that distinct pairing gaps are considered when investigating MFs in Sec.~\ref{Sec:Top_Su}.

Our goal is to recompute Eq.~(\ref{Eq:susceptibility}) with 
\begin{eqnarray}
\mathcal{H}_{\textrm{eff}} &=& \sum_{\nu,P} \int \frac{\hbar dr}{2\pi} \left\{
u_{\nu P} K_{\nu P} \left[ \triangledown \theta_{\nu P}(r) \right]^2 \right.
\nonumber\\ && \hspace{0.65in}
+  \left.
\frac{u_{\nu P}}{K_{\nu P}} \left[ \triangledown \phi_{\nu P}(r) \right]^2
\right\} \nonumber\\
&&+ \frac{ \Delta_{s}^2 }{\hbar v_{F}} \int \frac{dr}{2\pi} 
 \left( \phi_{cA}^2 + \phi_{sS}^2 + \theta_{cS}^2 + \theta_{sA}^2 \right).
\end{eqnarray}
Following Refs.~\citenum{Voit:1998,Starykh:1999,Meng:2013,Meng:2014}, we take the approximation $u_{\nu P} \approx v_{F}$, and find the zero-temperature correlation functions in the limits of 
$|\tilde{r}|\equiv \sqrt{r^2 + v_{F}^2\tau^2} \gg \frac{\hbar v_{F}}{\Delta_{s}}$ and $|\tilde{r}| \ll \frac{\hbar v_{F}}{\Delta_{s}}$.
In the $|\tilde{r}| \gg \frac{\hbar v_{F}}{\Delta_{s}}$ limit, we get
\begin{widetext}
\begin{eqnarray}
\chi_{AA}^{x}(r,\tau) &=& \chi_{AA}^{y}(r,\tau) \nonumber\\
 &=&  \frac{-\cos(2k_{F}r)}{(2\pi a)^2} 
\left[ \frac{a}{\sqrt{r^2 + (v_{F}|\tau|+a)^2}}\right]^{\frac{1}{2} (K_{cS} + \frac{1}{K_{sS}})}  
\left[ \frac{\Delta_{s} a }{ \hbar v_{F} }\right]^{\frac{1}{2}(K_{cA}+ \frac{1}{K_{sA}})} 
\exp{ \left[ -(C_{cS}+C_{sS}) \frac{\Delta_{s} |\tilde{r}|}{\hbar v_{F}}  \right]}, \\
\chi_{AA}^{z}(r,\tau)
 &=&  \frac{-\cos(2k_{F}r)}{(2\pi a)^2}
\left[ \frac{a}{\sqrt{r^2 + (v_{F}|\tau|+a)^2}}\right]^{\frac{1}{2} (K_{cS} + K_{sA})}  
\left[ \frac{\Delta_{s} a }{\hbar v_{F}}\right]^{\frac{1}{2}(K_{cA}+ K_{sS})} \exp{ \left[ -(C_{cS}+C_{sA}) \frac{\Delta_{s} |\tilde{r}|}{\hbar v_{F}}  \right]},
\end{eqnarray}
while in the limit of $|\tilde{r}| \ll \frac{\hbar v_{F}}{\Delta_{s}}$, we obtain
\begin{eqnarray}
\chi_{AA}^{x}(r,\tau) = \chi_{AA}^{y}(r,\tau)
 &=&  \frac{-\cos(2k_{F}r)}{(2\pi a)^2} 
\left[ \frac{a}{\sqrt{r^2 + (v_{F}|\tau|+a)^2}}\right]^{2 g_{x}}, \\
\chi_{AA}^{z}(r,\tau)
 &=&  \frac{-\cos(2k_{F}r)}{(2\pi a)^2}
\left[ \frac{a}{\sqrt{r^2 + (v_{F}|\tau|+a)^2}} \right]^{2 g_{z}},
\end{eqnarray}
\end{widetext}
where $C_{cS},C_{sS},C_{sA}$ are constants of order one. For large distance and long time, $|\tilde{r}| \gg \frac{\hbar v_{F}}{\Delta_{s}}$, which corresponds to small momenta and low frequencies, the correlation between the electrons is cut off by the superconducting gap, and the correlation functions exhibit an exponential decay. 
For small distance and short time, $|\tilde{r}| \ll \frac{\hbar v_{F}}{\Delta_{s}}$, in contrast, the correlation functions retain the gapless form. 
If the spin rotational symmetry is preserved, $K_{sS}=K_{sA}=1$, then the RKKY interaction is isotropic as expected. 

Fourier transforming into the momentum space and Matsubara frequency domain, and taking $q=2k_{F}$, $i\omega_{n} \rightarrow \omega + i\delta$, and $\omega \rightarrow 0$, we obtain the zero-temperature value of the peaks of the static spin susceptibility, 
\begin{eqnarray}
\chi_{AA}^{\mu}(q=2k_{F}, \omega \rightarrow 0)
&\approx& -\frac{1}{4\pi \hbar v_{F}} \frac{1}{2-2g_{\mu}} \nonumber \\
&& \times 
\left[ \left( \frac{\Delta_{a}}{\Delta_{s}} \right)^{(2-2g_{\mu})} -1\right],
\end{eqnarray} 
which gives Eq.~(\ref{Eq:chi_gapped}) in Sec.~\ref{SubSec:chi_gapped}.

\section{\label{Sec:MF_solution} MF solutions}

In this appendix, we first examine the bulk spectrum of $H_{\textrm{top}}^{(1)}$ and $H_{\textrm{top}}^{(2)}$ in Eq.~(\ref{Eq:H_top_block}), and then solve the Schr\"odinger equation for the MF solutions. The bulk spectrum of $H_{\textrm{top}}^{(1)}$, defined as $E^{(1)}(k)$, is too complicated to reproduce here. Instead, we express it as the roots of the following two quartic equations, 
\begin{widetext} 
\begin{eqnarray}
0&=& \left[ E^{(1)}(k)\right]^4-2 \left[ \left(\hbar v_{F}k \right)^2 + \Delta_{m}^2 + \Delta_{s,-}^2 + \Delta_{s,+}^2 + \Delta_{Z}^2\right] \left[ E^{(1)}(k) \right]^2 \pm 8 \Delta_{s,-} \Delta_{s,+} \Delta_{Z} E^{(1)}(k) \nonumber \\
&& + \left(\hbar v_{F} k\right)^4
+2 \left( \Delta_{m}^2 + \Delta_{s,-}^2 + \Delta_{s,+}^2 - \Delta_{Z}^2 \right)\left(\hbar v_{F} k\right)^2 
\nonumber \\
&& + 
\left[ \Delta_{Z} + \left( \Delta_{s,-} + \sqrt{\Delta_{s,+}^2 - \Delta_{m}^2} \right) \right] 
\left[ \Delta_{Z} - \left( \Delta_{s,-} + \sqrt{\Delta_{s,+}^2 - \Delta_{m}^2} \right) \right] \nonumber\\
&& \times \left[ \Delta_{Z} + \left( \Delta_{s,-} - \sqrt{\Delta_{s,+}^2 - \Delta_{m}^2} \right) \right] 
\left[ \Delta_{Z} - \left( \Delta_{s,-} - \sqrt{\Delta_{s,+}^2 - \Delta_{m}^2} \right) \right],
\label{Eq:quarticEk}
\end{eqnarray}
\end{widetext}
which differ by the sign of the term linear in $E^{(1)}(k)$. For nonzero $\Delta_{s,-} \Delta_{s,+} \Delta_{Z}$, each equation gives four roots, and in general there are eight non-degenerate energy bands from the $H_{\textrm{top}}^{(1)}$ block.
The bulk spectrum of $H_{\textrm{top}}^{(2)}$ is given by $\pm E_{e,\pm}^{(2)}(k)$ and $\pm E_{i,\pm}^{(2)}(k)$, where  
\begin{eqnarray}
E_{e,\pm}^{(2)}(k)&\equiv& \sqrt{ \left(\hbar v_{F} k\right)^2 + \left[ \Delta_{s}^{(e)} \right]^{2}} \pm \Delta_{Z}, \\
E_{i,\pm}^{(2)}(k)&\equiv& \sqrt{ \left(\hbar v_{F} k\right)^2 + \left[ \Delta_{s}^{(i)} \right]^{2}} \pm \Delta_{Z}.
\end{eqnarray}
In the absence of the superconductivity and the Zeeman field ($\Delta_{s}^{(e)}=\Delta_{s}^{(i)}=\Delta_{Z}=0$), half of the energy bands, $E^{(1)}(k)$, are gapped by the nuclear spin helix, whereas the other half, $E_{e/i,\pm}^{(2)}(k)$, remains gapless as discussed in Sec.~\ref{Sec:feedback}. 
At finite $\Delta_{s}^{(e)}\Delta_{s}^{(i)}\Delta_{Z}$, there exists a regime where the bulk spectrum has band touching points at $k=\pm k_{0}$, where
\begin{eqnarray}
k_{0} &\equiv& \frac{1}{ \hbar v_{F} } \left[ \Delta_{Z}^2 - \Delta_{m}^2 - \Delta_{s,-}^2 - \Delta_{s,+}^2 \right. \nonumber\\
&& \left. + 2 \sqrt{
\left(\Delta_{m}^2 + \Delta_{s,-}^2\right) \Delta_{s,+}^2 - \Delta_{m}^2 \Delta_{Z}^2 } \right]^{\frac{1}{2}} .
\end{eqnarray} 
This regime is given by the inequalities,
\begin{eqnarray}
&& k_{0}^2 > 0, \\
&& \left(\Delta_{m}^2 + \Delta_{s,-}^2\right) \Delta_{s,+}^2 - \Delta_{m}^2 \Delta_{Z}^2 > 0,
\end{eqnarray}
which is marked in blue color in Fig.~\ref{Fig:phase}. 
In this regime the system is a non-topological gapless superconductor, and therefore not of our interest. In other regimes, the system is fully gapped except for the following curves, where the bulk gap closes at $k=0$,
\begin{subequations} 
\begin{eqnarray}
C_{+}^{(1)}:&& \;  \left( \Delta_{Z} + \Delta_{s,-} \right)^2 + \Delta_{m}^2 - \Delta_{s,+}^2 = 0, \\
C_{-}^{(1)}:&& \; \left( \Delta_{Z} - \Delta_{s,-} \right)^2 + \Delta_{m}^2 - \Delta_{s,+}^2 = 0, \\
C_{e}^{(2)}: && \; \left| \Delta_{Z} \right| - \left| \Delta_{s,+} + \Delta_{s,-} \right| = 0, \\
C_{i}^{(2)}: && \; \left| \Delta_{Z} \right| - \left| \Delta_{s,+} - \Delta_{s,-} \right| = 0.
\end{eqnarray}
\end{subequations}
These gap closing curves are marked as $C_{+}^{(1)}$, $C_{-}^{(1)}$, $C_{e}^{(2)}$, and $C_{i}^{(2)}$, and plotted as black solid curves in Fig.~\ref{Fig:phase}.

Having the gap closing boundaries, we now discuss the topological properties of the Hamiltonian, and investigate the criterion for topological phases.~\cite{Klinovaja:2012a} To this end, we consider a semi-infinite nanotube with an open left end, and solve the Schr\"odinger equation at zero energy with the boundary condition of the MF wave function being zero at $r=0$. Since the boundary condition is imposed in the real space, we need to examine it in the basis, $\overline{\phi} \equiv$ $(c_{A,\uparrow}^{\dagger}, c_{A,\downarrow}^{\dagger},
c_{B,\uparrow}^{\dagger}, c_{B,\downarrow}^{\dagger},
c_{A,\uparrow}, c_{A,\downarrow},c_{B,\uparrow}, c_{B,\downarrow})$, which is related to the slowly varying fields $R_{\gamma,\sigma}$ and $L_{\gamma,\sigma}$ by Eqs.~(\ref{Eq:diag}) -- (\ref{Eq:transform}). 

We first focus on the Hamiltonian density of the first block, $H_{\textrm{top}}^{(1)}(r)$, and solve the Schr\"odinger equation, $H_{\textrm{top}}^{(1)}(r) \Phi^{(1)}_{\pm}(r) = 0$. 
We are looking for the localized states at the left end of the nanotube, so we use the ansatz
\begin{eqnarray}
\left[ \Phi^{(1)}_{\pm}(r) \right]^{T} &=&  e^{-\kappa_{1,\pm} r} \left( A_{1,\pm},B_{1,\pm},C_{1,\pm},D_{1,\pm}, \right. \nonumber\\
&& \hspace{0.5in} \left.
A_{1,\pm}^{*},B_{1,\pm}^{*},C_{1,\pm}^{*},D_{1,\pm}^{*} \right),
\label{Eq:Phi1_ansatz}
\end{eqnarray}
which incorporates the self-conjugate property of MFs. This gives the evanescent wave functions with the exponential decay determined by the $\kappa_{1,\pm}$ values,
\begin{eqnarray}
\kappa_{1,\pm} &\equiv& \frac{1}{ \hbar v_{F} }  \left[ \Delta_{m}^2 + \Delta_{s,-}^2 + \Delta_{s,+}^2 - \Delta_{Z}^2 \right. \nonumber\\
&& \left. \pm 2 \sqrt{
\left(\Delta_{m}^2 + \Delta_{s,-}^2\right) \Delta_{s,+}^2 - \Delta_{m}^2 \Delta_{Z}^2 } \right]^{\frac{1}{2}},
\end{eqnarray}
which have positive real parts and thus give normalizable wave functions when the bulk spectrum is fully gapped. 
These $\kappa_{1,\pm}$ values can also be obtained by setting $E^{(1)}(k)=0$ and $k=i \kappa_{1,\pm}$ in Eq.~(\ref{Eq:quarticEk}). 
After numerically solving the matrix eigenvalue equation, we find that each of the $\kappa_{1,\pm}$ values gives 
two eigenvectors of the form in Eq.~\eqref{Eq:Phi1_ansatz}, which result in four normalizable wave functions, denoted as
$\Phi^{(1)}_{+,1}(r)$, $\Phi^{(1)}_{+,2}(r)$, $\Phi^{(1)}_{-,1}(r)$, and $\Phi^{(1)}_{-,2}(r)$. 

Similarly, for the second block, 
$H_{\textrm{top}}^{(2)}(r)$, we find four zero-energy solutions for the Schr\"odinger equation, $H_{\textrm{top}}^{(2)}(r) \Phi^{(2)}_{e/i}(r) = 0$. These are simpler and can be obtained explicitly, as 
\begin{eqnarray}
\Phi^{(2)}_{e,1}(r) = 
\left(
\begin{array}{c}
1\\ 0\\ 0\\ F_{e}\\ 1\\ 0\\ 0\\ F_{e}^{*}
\end{array}
\right) e^{-\kappa_{2,e} r}, \,
\Phi^{(2)}_{e,2}(r) = 
\left(
\begin{array}{c}
i\\ 0\\ 0\\ -iF_{e}\\ -i\\ 0\\ 0\\ iF_{e}^{*}
\end{array}
\right) e^{-\kappa_{2,e} r}, \nonumber\\
\Phi^{(2)}_{i,1}(r) = 
\left(
\begin{array}{c}
0\\ 1\\ F_{i}\\ 0\\ 0\\ 1\\ F_{i}^{*}\\ 0
\end{array}
\right) e^{-\kappa_{2,i} r}, \,
\Phi^{(2)}_{i,2}(r)
= 
\left(
\begin{array}{c}
0\\ i\\ -iF_{i}\\ 0\\ 0\\ -i\\ iF_{i}^{*}\\ 0 
\end{array}
\right) e^{-\kappa_{2,i} r}, \nonumber\\
\end{eqnarray}
where 
\begin{subequations}
\begin{eqnarray}
F_{e/i} &\equiv& \frac{-i \sqrt{\left[\Delta_{s}^{(e/i)}\right]^2 -\Delta_{Z}^2} +\Delta_{Z}}{\Delta_{s}^{(e/i)}}, \\
\kappa_{2,e} &\equiv& \frac{1}{\hbar v_{F}} \sqrt{\left[\Delta_{s}^{(e)}\right]^2 -\Delta_{Z}^2}, \\
\kappa_{2,i} &\equiv& \frac{1}{\hbar v_{F}} \sqrt{\left[\Delta_{s}^{(i)}\right]^2 -\Delta_{Z}^2}.
\end{eqnarray}
\end{subequations}
Therefore, $\{\Phi^{(1)}_{\pm,n=1,2}(r), \Phi^{(2)}_{e/i,n=1,2}(r)\}$, which satisfy the Schr\"odinger equation and self-conjugation, form a set of eight basis wave functions. 
Using Eqs.~(\ref{Eq:diag}) -- (\ref{Eq:transform}), we obtain the corresponding wave functions, $\overline{\mathcal{B}} \equiv \{ \overline{\Phi}^{(1)}_{\pm,n=1,2}(r), \overline{\Phi}^{(2)}_{e/i,n=1,2}(r)\}$, in the $\overline{\phi}$ basis, and examine the boundary condition at the left end of the tube ($r=0$).
For a given set of system parameters, $(\Delta_{m},\Delta_{s,+},\Delta_{s,-},\Delta_{Z})$, the number of MFs, $N_{\textrm{MF}}$, is given as eight (the number of the column vectors in $\overline{\mathcal{B}}$) minus the number of the linearly independent vectors in $\overline{\mathcal{B}}$. 

The results are shown in Fig.~\ref{Fig:phase}. The gap closing curves, $C_{\pm}^{(1)}$ and $C_{e/i}^{(2)}$, correspond to the boundaries where $\kappa_{1,\pm}$ and $\kappa_{2,e/i}$ vanish, respectively. Moreover, the $\kappa_{1,\pm}$ values change from real or pure imaginary to complex numbers (but do not vanish) across the black dashed curve, marked as $C^{(3)}$,  
\begin{eqnarray}
C^{(3)}: && \; \left(\Delta_{m}^2 + \Delta_{s,-}^2\right) \Delta_{s,+}^2 - \Delta_{m}^2 \Delta_{Z}^2 = 0.
\end{eqnarray}
The yellow shaded region corresponds to a topological regime. The MF wave functions are linear superpositions of $\overline{\Phi}^{(1)}_{\pm,n=1,2}(r)$ and $\overline{\Phi}^{(2)}_{e/i,n=1,2}(r)$, and thus display multiple decay length scales, arising from $\kappa_{1,\pm}$ and $\kappa_{2,e/i}$.~\cite{Klinovaja:2012a} As a result, the localization length, $\xi_{\textrm{loc}}$, is determined by the largest length scale of the inverses of $\kappa_{1,\pm}$ and $\kappa_{2,e/i}$ or, equivalently, the smaller of $\Delta_{s}^{(e)}$ and $\Delta_{s}^{(i)}$.

\bibliographystyle{apsrev4-1}
\bibliography{MF_CNT}

\end{document}